\documentclass[preprint,prd,aps,nofootinbib]{revtex4}
\usepackage{graphics}

\begin{document}

\def\lvec#1{\vbox{\ialign{##\crcr$\leftarrow$\crcr\noalign{
 \kern-1pt\nointerlineskip}$\hfil\displaystyle{#1}\hfil$\crcr}}}
\def\rvec#1{\vbox{\ialign{##\crcr$\rightarrow$\crcr\noalign{
 \kern-1pt\nointerlineskip}$\hfil\displaystyle{#1}\hfil$\crcr}}}
\def\cstok#1{\leavevmode\thinspace\hbox{\vrule\vtop{\vbox{\hrule\kern1pt
\hbox{\vphantom{\tt/}\thinspace{\tt#1}\thinspace}}
\kern1pt\hrule}\vrule}\thinspace}

\begin{center}
\bibliographystyle{article}
{\Large \textsc{An introduction to quantum gravity}}
\end{center}

\author{Bryce S. DeWitt$^{\dagger}$ and Giampiero Esposito$^{1}$ \thanks{
Electronic address: giampiero.esposito@na.infn.it}} 

\affiliation{
${\ }^{1}$Istituto Nazionale di Fisica Nucleare, Sezione di Napoli,\\
Complesso Universitario di Monte S. Angelo, Via Cintia, Edificio 6, 80126
Napoli, Italy}

\vspace{0.4cm}
\date{\today}

\begin{abstract}
After an overview of the physical motivations for studying
quantum gravity, we reprint {\bf THE FORMAL STRUCTURE OF
QUANTUM GRAVITY}, i.e. the 1978 Carg\`ese Lectures by Professor B.S. 
DeWitt, with kind permission of Springer. The reader is therefore
introduced, in a pedagogical way, to the functional integral quantization
of gravitation and Yang--Mills theory. It is hoped 
that such a paper will remain useful
for all lecturers or Ph.D. students who face the task of introducing (resp.
learning) some basic concepts in quantum gravity in a relatively short time.
In the second part, we outline selected topics such as the braneworld 
picture with the same covariant formalism of the first part,
and spectral asymptotics of Euclidean quantum gravity with 
diffeomorphism-invariant boundary conditions. The latter might have
implications for singularity avoidance in quantum cosmology.
\end{abstract}

\maketitle
\bigskip
\vspace{2cm}

\section{Motivations for and approaches to quantum gravity}

The aim of theoretical physics is to provide a clear conceptual framework
for the wide variety of natural phenomena, so that not only are we able to
make accurate predictions to be checked against observations, but the 
underlying mathematical structures of the world we live in can also become
sufficiently well understood by the scientific community. What are therefore
the key elements of a mathematical description of the physical world? Can
we derive all basic equations of theoretical physics from a set of symmetry
principles? What do they tell us about the origin and evolution of the
universe? Why is gravitation so peculiar with respect to all other 
fundamental interactions?

The above questions have received careful consideration over the last decades,
and have led, in particular, to several approaches to a theory aimed at
achieving a synthesis of quantum physics on the one hand, and general
relativity on the other hand. This remains, possibly, the most important
task of theoretical physics. The need for a quantum theory of gravity is
already suggested from singularity theorems in classical cosmology. Such
theorems prove that the Einstein theory of general relativity leads to the
occurrence of spacetime singularities in a generic 
way \cite{PRSLA-A314-529}.

At first sight one might be tempted to conclude that a breakdown of all
physical laws occurred in the past, or that general relativity is severely
incomplete, being unable to predict what came out of a singularity. It has
been therefore pointed out that all these pathological features result from 
the attempt of using the Einstein theory well beyond its limit of validity,
i.e. at energy scales where the fundamental theory is definitely more
involved. General relativity might be therefore viewed as a low-energy limit
of a richer theory, which achieves the synthesis of both the
{\bf basic principles} of modern physics and the 
{\bf fundamental interactions} in the form currently known.
So far, no less than 16 major approaches to quantum gravity have been
proposed in the literature. In alphabetical order (to avoid being
affected by our own preference) they are as follows.
\vskip 0.3cm
\noindent
1. Affine quantum gravity \cite{gr-qc/0612168}
\vskip 0.3cm
\noindent
2. Asymptotic quantization \cite{APNYA-62-582, PRLTA-46-573}.
\vskip 0.3cm
\noindent
3. Canonical quantum gravity \cite{PHRVA-160-1113, 
CQGRD-1-621, CQGRD-1-633, PRLTA-57-2244, NUCIA-B110-1137}
\vskip 0.3cm
\noindent
4. Condensed-matter view: the universe in a helium droplet \cite{IMPHA-117-1}.
\vskip 0.3cm
\noindent
5. Manifestly covariant quantization \cite{RMPHA-29-497, PHRVA-162-1195, 
AHPAA-A20-69, NUPHA-B266-709, NUPHA-B355-712, CQGRD-9-895, NUPHA-B731-389}.
\vskip 0.3cm
\noindent
6. Euclidean quantum gravity \cite{PHRVA-D15-2752, PHRVA-D18-1747}.
\vskip 0.3cm
\noindent
7. Lattice formulation \cite{PHRVA-D35-1194, CQGRD-9-595}.
\vskip 0.3cm
\noindent
8. Loop space representation \cite{PRLTA-61-1155, NUPHA-B331-80}.
\vskip 0.3cm
\noindent
9. Non-commutative geometry \cite{hep-th/0206006}.
\vskip 0.3cm
\noindent
10. Quantum topology \cite{CQGRD-6-1509}, motivated by Wheeler's
quantum geometrodynamics \cite{APNYA-2-604}.
\vskip 0.3cm
\noindent
11. Renormalization group and Weinberg's asymptotic 
safety \cite{PHRVA-D57-971, hep-th/0511260}.
\vskip 0.3cm
\noindent
12. R-squared gravity \cite{PHRVA-D16-953}.
\vskip 0.3cm
\noindent
13. String and brane theory \cite{hep-th/0212247, NJOPF-7-201, 
PHRVA-D74-084033}.
\vskip 0.3cm
\noindent
14. Supergravity \cite{PHRVA-D13-3214, PRPLC-68-189}.
\vskip 0.3cm
\noindent
15. Triangulations \cite{gr-qc/9812080, NUPHA-B542-349, arXiv:0711.0273}
and null-strut calculus \cite{CQGRD-6-659}.
\vskip 0.3cm
\noindent
16. Twistor theory \cite{PRPLC-6-241, CQGRD-16-A113}. 

After such a broad list of ideas, we hereafter focus on what can be
taught in a Ph.D. course aimed at students with a field-theoretic
background. We have therefore chosen to reprint, from section 2
to section 17, the 1978 DeWitt Lectures at Carg\`ese, with kind 
permission of Springer and of Professor C. DeWitt--Morette. 
In the second part we select
two related topics, i.e. the braneworld picture and the spectral
asymptotics of Euclidean quantum gravity, since they have possibly
a deep impact on the current attempts to describe a quantum origin
of the physical universe. Relevant group-theoretical material is
summarized in the Appendix.

\section{Introduction and notation}

In 1956 Utiyama pointed out that the gravitational field can be regarded as 
a non-Abelian gauge field. In 1963 Feynman found that in order to construct 
a quantum perturbation theory for a non-Abelian gauge field he had to introduce new graphical rules not previously encountered in quantum field theory. He
showed, in one-loop order, that to preserve unitarity one must add to every 
standard closed-loop graph another, involving a closed integral-spin fermion
loop. In 1966 an explicitly gauge invariant functional-integral algorithm 
was found which extended Feynman's new rules to all orders (DeWitt (1967b)).
A short time later it was shown that the algorithm could be obtained by a
method of factoring out the gauge group (Fadde'ev and Popov (1967)).

Formally the gravitational field and the Yang--Mills field can be treated
identically. In the computation of amplitudes for specific physical
processes, however, the two differ by the fact that the Yang--Mills field
yields a renormalizable theory while the gravitational field does not.

Some of the proposals that have been made for dealing with quantum gravity
despite its nonrenormalizability will be discussed briefly later. But it must 
be admitted at the outset that we are dealing with an incomplete theory. 
The student may take comfort in the fact that every formal statement will be
true for all field theories, even those, like supergravity, possessing
supergauge groups, provided they are formulated in such a way that the action
of the (super)gauge group on the field variables is expressible without use
of field equations, and the group operations thus given are closed.  

To emphasize the generality of the formalism we shall, most of the time, 
suppress field symbols such as $g_{\mu \nu}$ for the gravitational field
or $A_{\; \mu}^{\alpha}$ for the Yang-Mills field. The index $i$ (or
$j,k,l,$ etc.) will be understood to label not only a field component but
also a spacetime point $x$. Thus, in the gravitational case, $i$ will be
understood to stand for the set $\left \{ \mu, \nu, x \right \}$ and, in
the Yang-Mills case, for the set $\left \{ \alpha, \mu, x \right \}$. In a
supergauge theory the set $i$ may include spinor indices. When it does,
$i$ (or $\varphi^{i}$) is said to be fermionic; otherwise it is bosonic.

The reason for including the continuous label $x$ in the set $i$ is that
much of the formalism of quantum field theory is purely combinatorial, with
summation over dummy indices being accompanied by integration over spacetime.
In order to avoid having to write a lot of integral signs we lump $x$ and
the field indices together and adopt the convention that the repetition 
of a lower case Latin index implies a combined summation-integration. 
Correspondingly, a comma followed by a lower case Latin index will denote
{\bf functional} differentiation:
\begin{equation}
A_{,i} \equiv A {{\lvec \delta}\over \delta \varphi^{i}}.
\label{(1)}
\end{equation}
The change in a functional $A$ (of the fields $\varphi^{i}$) resulting
from an infinitesimal variation $\delta \varphi^{i}$ is then
\begin{equation}
\delta A=A_{,i}\delta \varphi^{i}.
\label{(2)}
\end{equation}

If $A$, or any of the $\varphi^{i}$, is fermionic one must distinguish left
from right differentiation:
$$
{ }_{i,}A \equiv {{\rvec \delta}\over \delta \varphi^{i}}A,
$$
$$
\delta A = \delta \varphi^{i} \; { }_{i,}A.
$$
Evidently
\begin{equation}
{ }_{i,}A=(-1)^{i(A+1)}A_{,i}
\label{(3)}
\end{equation}
where we adopt the rule that when an index (such as $i$) or a dynamical
quantity (such as $A$) appears as an exponent of $-1$ it is to be
understood as assuming the value $0$ or $1$ according as it is bosonic
or fermionic. The summation-integration convention is not to be understood 
as applying to indices appearing as exponents. Such indices may participate 
in summation-integrations induced by their appearance twice elsewhere
in an expression, but they temselves may not induce 
summation-integrations. Note that we are here treating all quantities
($A,\varphi^{i},\delta \varphi^{i}$ etc.) as {\bf supernumbers}, i.e.
as even or odd elements of an infinite-dimensional Grassmann algebra.
Bosonic quantities commute with everything; fermionic quantities 
anticommute among themselves. In the quantum theory this perfect
commutativity or anticommutativity is broken. The corresponding quantities
will then be written in boldface.

The following notation will sometimes be convenient for expressing
repeated functional differentiation:
\begin{equation}
{ }_{...ij,}A_{,kl...} \equiv
{{\rvec \delta}\over \delta \varphi^{i}}
{{\rvec \delta}\over \delta \varphi^{j}} A 
{{\lvec \delta}\over \delta \varphi^{k}}
{{\lvec \delta}\over \delta \varphi^{l}} ...
\label{(4)}
\end{equation}
Note the particular examples:
\begin{equation}
\varphi_{\; ,j}^{i} = \delta_{\; j}^{i},
\label{(5)}
\end{equation}
\begin{equation}
(\varphi^{i}\varphi^{j})_{,kl}=(-1)^{ij}\delta_{\; k}^{i}
\; \delta_{\; l}^{j}+\delta_{\; l}^{i} \; \delta_{\; k}^{j}.
\label{(6)}
\end{equation}
If $x$ belongs to the set $i$ and $x'$ belongs to the set $j$, the
``generalized Kronecker delta'' $\delta_{\; j}^{i}$ includes, as a
factor, the spacetime delta function $\delta(x,x')$.

When we are displaying specific details of a given field theory lower 
case indices from the middle of the Greek alphabet will be used to label
tensor components. Coordinates in a given local patch, or chart, will be
denoted by $x^{\mu}$, with $\mu$ running from $0$ to $n-1$, $n$ being the
dimensionality of spacetime. (With an eye to the ultimate application
of methods such as dimensional regularization and the renormalization
group, we do not here hold $n$ fixed at $4$.) Commas followed by lower-case
mid-alphabet Greek indices will denote ordinary differentiation with 
respect to the coordinates.

The following abbreviations will be useful:
\begin{equation}
{\delta g_{\mu \nu} \over \delta g_{\sigma' \tau'}}
=\delta_{\mu \nu}^{\; \; \; \sigma' \tau'}
\equiv {1\over 2} \Bigr(\delta_{\mu}^{\; \sigma} \delta_{\nu}^{\; \tau}
+\delta_{\mu}^{\; \tau} \delta_{\nu}^{\; \sigma}\Bigr) \delta(x,x'),
\label{(7)}
\end{equation}
\begin{equation}
{\delta A_{\; \mu}^{\alpha} \over \delta A_{\; \nu'}^{\beta'}}
=\delta_{\; \mu \beta'}^{\alpha \; \; \; \nu'} \equiv
\delta_{\; \beta}^{\alpha} \delta_{\mu}^{\; \nu}\delta(x,x').
\label{(8)}
\end{equation}
It is straightforward to verify that
\begin{equation}
\delta_{\mu \nu \; \; \; \; \; \; ;}^{\; \; \; \sigma' \tau' \; \nu}
=-\delta_{\mu \; \; ;}^{\; \sigma' \; \tau'}
-\delta_{\mu \; \; ;}^{\; \tau' \; \sigma'},
\label{(9)}
\end{equation}
\begin{equation}
\delta_{\; \mu \beta' \; \; ;}^{\alpha \; \; \; \nu' \; \mu}
=-\delta_{\; \beta' ;}^{\alpha \; \; \; \nu'},
\label{(10)}
\end{equation}
where
$$
\delta_{\mu}^{\; \sigma'} \equiv \delta_{\mu}^{\; \sigma} \delta(x,x'), \;
\delta_{\; \beta'}^{\alpha} \equiv \delta_{\; \beta}^{\alpha} \delta(x,x'),
$$
the semicolons denote covariant differentiation, and tensor indices may be
lowered and raised by the metric tensor $g_{\mu \nu}$ and its inverse
$g^{\mu \nu}$ respectively. (Note that we always leave the semicolons in
the lower position regardless of what happens to the indices). The delta
functions appearing above are 2-point tensor densities, or 
{\bf bitensor densities}, of total weight unity. In eqs. (9) and (10) the
apportionment of the weight between the points $x$ and $x'$ is arbitrary;
in eqs. (7) and (8) all the weight is at $x'$. In eqs. (9) and (10) the
derivatives on the left are at $x$, on the right at $x'$.

In eqs. (8) and (10) lower case Greek indices from the beginning of the
alphabet appear. These are associated with the Yang-Mills group. The laws
of covariant differentiation of tensors bearing various kinds of indices
are determined as follows: let the symbol $T$ represent a tensor field,
with indices suppressed, in which we imagine all the components strung out
in a single column. Let $T$ also be coupled to the Yang-Mills field. Then
\begin{equation}
T_{;\mu} \equiv T_{,\mu}+G_{\; \sigma}^{\nu}\Gamma_{\; \nu \mu}^{\sigma}T
+G_{\alpha}A_{\; \mu}^{\alpha}T,
\label{(11)}
\end{equation}
where
\begin{equation}
\Gamma_{\; \nu \mu}^{\sigma} \equiv {1\over 2}g^{\sigma \tau}
(g_{\tau \nu,\mu}+g_{\tau \mu,\nu}-g_{\nu \mu,\tau}),
\label{(12)}
\end{equation}
and the $G_{\; \nu}^{\mu}$ and $G_{\alpha}$ are respectively the matrix
generators of the representations of the linear group and Yang-Mills Lie
group to which $T$ corresponds. These generators satisfy
\begin{equation}
\Bigr[G_{\; \nu}^{\mu},G_{\; \tau}^{\sigma}\Bigr]
=\delta_{\; \tau}^{\mu}G_{\; \nu}^{\sigma}
-\delta_{\; \nu}^{\sigma} G_{\; \tau}^{\mu},
\label{(13)}
\end{equation}
\begin{equation}
[G_{\alpha},G_{\beta}]=G_{\gamma}f_{\; \alpha \beta}^{\gamma},
\label{(14)}
\end{equation}
where $f_{\; \alpha \beta}^{\gamma}$ are the structure constants of the
Yang-Mills Lie group. When the suppressed indices on $T$ are restored their
positions generally determine the particular representations involved.
Thus a Yang-Mills index in the upper position indicates the adjoint
representation of the Lie group and one in the lower position the 
contragredient representation, etc. For physical reasons (positive
probability) the Yang-Mills Lie group is required to be compact. Therefore
given representations and their contragredient forms are equivalent, and
Yang-Mills indices may be lowered (and raised) by the matrix 
$\gamma_{\alpha \beta}$ (and its inverse $\gamma^{\alpha \beta}$) that
connects the adjoint and co-adjoint representations. When the Lie group
is simple $\gamma_{\alpha \beta}$ may be taken to be the Kronecker delta,
and all Yang-Mills indices may be dropped to the lower position.

We make no attempt here to list the rather complicated additional
structures that appear in supergravity theories. (The student should
consult the published literature for those details). We remark only
that when spinors are present the {\bf local frame group} and its spin
representations must be introduced. The local frame group is completely
analogous to the Yang-Mills group and makes a corresponding contribution
to the covariant derivative on the right side of eq. (11), with
$A_{\; \mu}^{\alpha}$ replaced by the connection components in the local
frame and the $G_{\alpha}$ replaced by the generators of the relevant
spin representation of the Lorentz group. (See De Witt (1965) for 
details).

In the next section we shall show how to place the Yang-Mills group and
the diffeomorphism group (with which tensor indices are associated) on a
common footing. Despite the analogies between the two groups, as displayed
for example by the similarity of the second and third terms on the
right of eq. (11), the diffeomorphism group is much more complicated
than the Yang-Mills group and much less is known about its structure. 
Moreover, there is a lack of symmetry between the two groups in the fact
that when they are combined into a single group (as is necessary when both
Yang-Mills and gravitational fields are present) they are united not in
a direct product but in a semi-direct product based on the automorphisms 
of the Yang-Mills group under diffeomorphisms. The same is true for the
combined diffeomorphism and local frame groups, when spinor fields are
present.

Our list of notational conventions is completed with the following
statements and definitions:
\begin{equation}
T_{; \mu \nu}-T_{; \nu \mu}=-\Bigr(G_{\; \tau}^{\sigma} 
R_{\; \sigma \mu \nu}^{\tau}+G_{\alpha}F_{\; \mu \nu}^{\alpha}\Bigr)T ,
\label{(15)}
\end{equation}
\begin{equation}
R_{\; \sigma \mu \nu}^{\tau} \equiv
\Gamma_{\; \sigma \nu, \mu}^{\tau}
-\Gamma_{\; \sigma \mu,\nu}^{\tau}
+\Gamma_{\; \mu \rho}^{\tau} \Gamma_{\; \sigma \nu}^{\rho}
-\Gamma_{\; \nu \rho}^{\tau} \Gamma_{\; \sigma \mu}^{\rho},
\label{(16)}
\end{equation}
\begin{equation}
F_{\; \mu \nu}^{\alpha} \equiv A_{\; \nu , \mu}^{\alpha}
-A_{\; \mu , \nu}^{\alpha}+f_{\; \beta \gamma}^{\alpha}
A_{\; \mu}^{\beta} A_{\; \nu}^{\gamma},
\label{(17)}
\end{equation}
\begin{equation}
R_{\mu \nu} \equiv R_{\; \mu \sigma \nu}^{\sigma}, \;
R \equiv R_{\mu}^{\; \mu}.
\label{(18)}
\end{equation}
Unless otherwise specified we shall assume that spacetime is globally
hyperbolic and complete\footnote{There is some evidence (although hardly
overwhelming yet) that quantization suppresses the singularities in
spacetime that often develop automatically in classical general relativity.}.
Without loss of generality $x^{0}$ may then be assumed to be a global
time coordinate, in the sense that it defines a foliation of spacetime
into smooth complete hypersurfaces $x^{0}={\rm constant}$, arranged in a
temporal order. These hypersurfaces need not be everywhere spacelike,
although if they are noncompact they must be asymptotically spacelike.
The signature of the metric tensor will be $-+++$ ..., and units (when
needed) will be chosen to be ``absolute'', with ${\hbar}=c=32\pi G=1$. The
absolute units of length, time and mass respectively are 
$1.6 \times 10^{-32}{\rm cm}$., $5 \times 10^{-43}$ sec. and
$2 \times 10^{-6}$ g., which gives an idea of the domains in which quantum
gravity becomes relevant.

\section{The gauge group}

The gauge group of quantum gravity is the diffeomorphism group and that
of Yang-Mills theory is the Yang-Mills group. We begin with the latter.

Elements of the Yang-Mills group are locally parametrized by a set of
differentiable scalar functions $\xi^{\alpha}(x)$, with 
$\xi^{\alpha}=0$ denoting the identity element. Elements infinitesimally
close to the identity are parameterized by infinitesimal scalars. The 
action of such an element on the Yang-Mills potentials is given by
\begin{equation}
\delta A_{\; \mu}^{\alpha}=-\delta \xi_{\; , \mu}^{\alpha}
+f_{\; \gamma \beta}^{\alpha}A_{\; \mu}^{\beta} \delta \xi^{\gamma}
=-\delta \xi_{\; ;\mu}^{\alpha},
\label{(19)}
\end{equation}
the covariant derivative being determined by noting that 
$\delta \xi^{\alpha}$ transforms (under inner automorphisms) according to
the adjoint representation of the group. It will be convenient to
rewrite eq. (19) in the form
\begin{equation}
\delta A_{\; \mu}^{\alpha}=\int Q_{\; \mu \beta'}^{\alpha}
\delta \xi^{\beta'} d^{n}x', \;
d^{n}x' \equiv \prod_{\mu=0}^{n-1}dx^{\mu},
\label{(20)}
\end{equation}
or, in the generic notation,
\begin{equation}
\delta \varphi^{i}=Q_{\; \alpha}^{i} \delta \xi^{\alpha},
\label{(21)}
\end{equation}
where
\begin{equation}
Q_{\; \mu \beta'}^{\alpha} \equiv -\delta_{\; \beta' ; \mu}^{\alpha}.
\label{(22)}
\end{equation}
In passing from eq. (20) to the generic form (21) one replaces the labels
$\alpha,\mu,x$ by the index $i$ and the labels $\beta',x'$ by the index
$\alpha$, and one understands that repetition of the latter index
implies a combined summation-integration.

In quantum gravity the action of the diffeomorphism group can be 
expressed in identical generic form. The diffeomorphism group is the
group of mappings $f: M \rightarrow M$ of the spacetime manifold $M$
into itself such that $f$ is one-to-one and both $f$ and $f^{-1}$
are differentiable. In practice one may require $f$ and $f^{-1}$
to be $C^{\infty}$ and, if the sections $x^{0}={\rm constant}$ are
noncompact, to reduce asymptotically (i.e. ``at spatial infinity'')
to the local identity mapping. Such mappings define a ``dragging'' of 
all tensor fields defined on $M$, and if the mapping is infinitesimally
close to the identity the ``dragging'' may be viewed as a physical
displacement of the fields through an infinitesimal vector $\delta \xi$.
If all fields, including the metric (i.e. gravitational) field, are
displaced by the same amount the physics remains unchanged. It is
conventional in physics, therefore, to adopt an opposite viewpoint 
and to regard an infinitesimal diffeomorphism as leaving the ``physical''
points of the manifold untouched while dragging all coordinate patches
(i.e. the complete atlas) through the negative vector $\delta \xi$. Locally
this is expressed by the coordinate transformation
$x^{\mu} \rightarrow \xi^{\mu}$ where
\begin{equation}
\xi^{\mu}=x^{\mu}+\delta \xi^{\mu}.
\label{(23)}
\end{equation}

Let $T$ be a tensor field and $\delta T$ its change under dragging 
through $\delta \xi$. The Lie derivative of $T$ with respect to
$\delta \xi$ is defined by
\begin{equation}
{\cal L}_{\delta \xi}T=-\delta T.
\label{(24)}
\end{equation}
Let $p$ be a point of $M$ and $p'$ the point to which it is dragged
under $\delta \xi$. Then the coordinates of $p$ in the new coordinate
system (23) are identical with those of $p'$ in the old coordinate
system. Moreover, the components of $T$ at $p$ in the new coordinate 
system are identical with those of $T+\delta T$ at $p'$ in the old
coordinate system. One has only to cast eq. (24) into component
language, therefore, to regard the Lie derivative as expressing the
negative of the change in the {\bf functional form} of the components
of $T$, {\bf viewed as functions of the local coordinates}, under the
diffeomorphism. This enables one to compute
\begin{eqnarray}
{\cal L}_{\delta \xi}T &=& T_{,\mu}\delta \xi^{\mu}
-G_{\; \mu}^{\nu}T \delta \xi_{\; ,\nu}^{\mu} \nonumber \\
&=& T_{;\mu}\delta \xi^{\mu}-G_{\; \mu}^{\nu}T 
\delta \xi_{\; ;\nu}^{\mu},
\label{(25)}
\end{eqnarray}
which yields, in particular, the gauge transformation law for the
metric tensor:
\begin{equation}
\delta g_{\mu \nu}=-\Bigr({\cal L}_{\delta \xi}g \Bigr)_{\mu \nu}
=-\delta \xi_{\mu ; \nu}-\delta \xi_{\nu ; \mu}.
\label{(26)}
\end{equation}
Equation (26) may be rewritten
\begin{equation}
\delta g_{\mu \nu}=\int Q_{\mu \nu \sigma'}\delta \xi^{\sigma'}d^{n}x', 
\label{(27)}
\end{equation}
\begin{equation}
Q_{\mu \nu \sigma'} \equiv -\delta_{\mu \sigma'; \nu}
-\delta_{\nu \sigma'; \mu},
\label{(28)}
\end{equation}
which, if the labels $\mu,\nu,x$ are replaced by $i$ and the labels
$\sigma',x'$ by $\alpha$, takes the generic form (21).

Lower case Greek indices from the first part of the alphabet, as in
eq. (21), will from now on be called {\bf group} indices. If the group
indices are allowed to label fermionic as well as bosonic gauge parameters
then the generic form (21) holds also for the {\bf super}gauge 
transformations of supergravity theories. Although we shall not go into
the specific details of such theories we shall, in all that follows,
allow for their possible presence.

\section{Structure constants}

By invoking the requirement that the commutator of two infinitesimal
gauge group operations be itself a group operation (the closure
property) one arrives at the functional differential identity
\begin{equation}
Q_{\; \alpha,j}^{i} Q_{\; \beta}^{j}-(-1)^{\alpha \beta}
Q_{\; \beta,j}^{i} Q_{\; \alpha}^{j}
=Q_{\; \gamma}^{i}C_{\; \alpha \beta}^{\gamma},
\label{(29)}
\end{equation}
where the $C$'s are certain coefficients known as the 
{\bf structure constants} of the gauge group. They possess the symmetry
\begin{equation}
C_{\; \alpha \beta}^{\gamma}=-(-1)^{\alpha \beta}
C_{\; \beta \alpha}^{\gamma}.
\label{(30)}
\end{equation}

The structure constants of the Yang-Mills group may be determined by
straightforward computation from Eq. (19), (20) and (22). They are the
components of the following 3-point tensor density:
\begin{equation}
C_{\; \beta' \gamma''}^{\alpha}=f_{\; \; \beta \gamma}^{\alpha}
\delta(x,x') \delta(x,x'').
\label{(31)}
\end{equation}
The weights are at $x'$ and $x''$.

The structure constants of the diffeomorphism group may be determined
by recalling the commutation law for the Lie derivative:
\begin{equation}
\Bigr[{\cal L}_{X},{\cal L}_{Y}\Bigr]T={\cal L}_{[X,Y]}T.
\label{(32)}
\end{equation}
Here $[X,Y]$ is the Lie bracket of the vectors $X$ and $Y$:
\begin{equation}
[X,Y]={\cal L}_{X}Y=-{\cal L}_{Y}X.
\label{(33)}
\end{equation}
The structure constants are the components of the 3-point tensor density
defined by
\begin{eqnarray}
\; & \; & \int d^{n}x' \int d^{n}x'' C_{\; \nu' \sigma''}^{\mu}
X^{\nu'} Y^{\sigma''}=-[X,Y]^{\mu} \nonumber \\
&=& X_{\; ,\nu}^{\mu} Y^{\nu}-Y_{\; ,\nu}^{\mu}X^{\nu}
=X_{\; ;\nu}^{\mu} Y^{\nu}-Y_{\; ;\nu}^{\mu}X^{\nu}.
\label{(34)}
\end{eqnarray}
Evidently
\begin{equation}
C_{\; \nu' \sigma''}^{\mu}=\delta_{\; \nu',\tau}^{\mu}
\delta_{\; \sigma''}^{\tau}-\delta_{\; \sigma'',\tau}^{\mu}
\delta_{\; \nu'}^{\tau}
=\delta_{\; \nu';\tau}^{\mu} \delta_{\; \sigma''}^{\tau}
-\delta_{\; \sigma'';\tau}^{\mu} \delta_{\; \nu'}^{\tau}.
\label{(35)}
\end{equation}
The weights are at $x'$ and $x''$.

The action of the gauge group on the field variables $\varphi^{i}$,
expressed by eq. (21), is a {\bf realization} of the group. This
realization is always a {\bf faithful} one, which implies that
$Q_{\; \alpha}^{i}X^{\alpha}=0$ for all $i$ if and only if 
$X^{\alpha}=0$ for all $\alpha$. By functionally differentiating eq.
(29) with respect to $\varphi^{k}$, multiplying by $Q_{\; \gamma}^{k}$,
judiciously permuting the indices $\alpha,\beta,\gamma$, adding the 
results, and invoking the faithfulness of the realization, one obtains
the following cyclic identity satisfied by the structure constants:
\begin{equation}
C_{\; \alpha \epsilon}^{\delta}C_{\; \beta \gamma}^{\epsilon}
+(-1)^{\alpha(\beta+\gamma)}C_{\; \beta \epsilon}^{\delta}
C_{\; \gamma \alpha}^{\epsilon}
+(-1)^{\gamma(\alpha+\beta)}C_{\; \gamma \epsilon}^{\delta}
C_{\; \alpha \beta}^{\epsilon}=0.
\label{(36)}
\end{equation}
In the case of the Yang-Mills group this identity reduces to the
corresponding identity for the constants $f_{\; \beta \gamma}^{\alpha}$.
In the case of the diffeomorphism group it is the Jacobi identity for
Lie brackets.

\section{Configuration space. Orbits}

For each point $x$ in the spacetime manifold $M$, the field $\varphi$
(index $i$ suppressed) takes its ``value'' in a certain 
finite-dimensional differentiable manifold $\Phi_{x}$, which may but
need not be a vector space or subspace thereof. In pure gravity
theory, for example, $\Phi_{x}$ is the subspace of
Sym$\Bigr(T_{x}^{\; *} \otimes T_{x}^{\; *}\Bigr)$ containing all local
symmetric covariant second rank tensors at $x$ having nonvanishing 
determinant and signature $-+++...$. Here $T_{x}^{\; *}$ is the dual
of the tangent space to $M$ at $x$, and ``Sym'' denotes the symmetric
part of the tensor product $T_{x}^{\; *} \otimes T_{x}^{\; *}$.

The set of all $\Phi_{x}$ with $x$ in $M$, may be regarded as forming
a fiber bundle over $M$. Each $\Phi_{x}$ is a fiber, and each field
$\varphi$ is a cross section of the bundle\footnote{If the fiber bundle
admits no global cross sections the field must be defined by introducing
overlapping patches.}. The bundle may but need not be a simple product
bundle.

The set of all cross sections, i.e. of all field configurations $\varphi$
may be assembled into a space $\Phi$ called the {\bf configuration space}.
Because of differentiability requirements on the field configurations
$\Phi$ is endowed naturally with a functional differentiable structure 
and may be viewed as an infinite dimensional differentiable manifold, or,
if $\varphi^{i}$ includes fermion fields, as a differentiable
{\bf super}manifold (also known as a $Z_{2}$-graded manifold. See
Kostant (1977)). Since $M$ is never compact (at least in the time
direction) the fields $\varphi$ are usually constrained to obey also
special boundary conditions ``at infinity''. In both Yang-Mills and gravity
theory these can be of considerable importance.

Let the gauge group be denoted by $G$ and let $\xi$ be an element of $G$.
Denote by ${ }^{\xi}\varphi$ the ``point'' of $\Phi$ to which $\varphi$ is
displaced under the action of $\xi$. The set of points 
${ }^{\xi}\varphi$ for all $\xi$ in $G$ is known as the {\bf orbit}
of $\varphi$ and denoted by Orb.($\varphi$). The set of all orbits can be
assembled into a space called the {\bf space of orbits}, denoted by the
quotient symbol $\Phi / G$. Since all fields on a given orbit describe the
same physics it is the space of orbits that constitutes the real
{\bf physical configuration space} of the theory. In pure gravity theory
$\Phi$ is the space, Lor(M), of Lorentzian (also called
pseudo-Riemannian) metrics on $M$, $G$ is the group, Diff(M), of 
diffeomorphisms of $M$, and the physical configuration space,
Lor(M)/Diff(M), is the space of {\bf Lorentzian geometries} on M.

Because gauge groups can be ``coordinatized'' by differentiable 
functions (i.e. the gauge parameters) $G$, like $\Phi$, can be regarded
as an infinite dimensional differentiable manifold (or supermanifold). In
Yang-Mills theory $G$ may have a simple product structure inherited from
the associated Lie group of the theory (see (31)), or it may itself
be a twisted bundle. The diffeomorphism group of gravity theory, by
contrast, cannot be viewed as a bundle but has a structure that is much
less well understood. Some, but only a little, of its complexity will
emerge as we go along.

Since both $\Phi$ and $G$ are differentiable (super) manifolds the
quotient space $\Phi / G$ too is a differentiable (super) manifold,
or rather it is a differentiable (super) manifold that may have a boundary.

To see how a boundary can arise consider a typical, i.e. {\bf generic},
orbit. Modulo a possible discrete center it is a {\bf copy} of $G$,
because it provides a realization of $G$ and has the same dimensionality.
Not all orbits need have this dimensionality. There is often a class
of degenerate orbits having fewer dimensions. These are the orbits that
remain invariant under the action of nontrivial continuous subgroups 
of $G$. They are the boundary points of $\Phi / G$. To see this think of
$\Phi$ as being ${\bf R}^{3}$ and $G$ as being the group of rotations
about a fixed axis. The orbits are then circles perpendicular to and
centered on the axis, and the orbit space is a half-plane whose boundary
points correspond to the points on the axis, which remain invariant under
the group.

The greater the dimensionality of the subgroup that leaves a given orbit
invariant, the smaller the dimensionality of the orbit. Fischer (1970) has
shown that if the invariance group has only one dimension then the orbit
is an ordinary boundary point of $\Phi / G$. If the invariance group has
two dimensions then the orbit lies on a boundary of the boundary and so on.
The whole orbit manifold, with its boundary, and its boundaries of
boundaries, etc., is known as a {\bf stratified} manifold.

In gravity theory the boundary orbits are the symmetrical geometries, i.e.
those that possess Killing vectors. The boundary structure of 
Lor($M$)/Diff($M$) in general depends critically on $M$. Since there
exists no complete classification of $n$-dimensional manifolds 
($n >3$) that can possess globally hyperbolic metric tensors, there exists
also no complete classification of possible configuration spaces for the
gravitational field. On some spacetimes there may be {\bf no}
Lorentzian geometries possessing Killing vectors. Such spacetimes are
called {\bf wild}. If $M$ is wild Lor($M$)/Diff($M$) has no boundary
points. For technical reasons it is frequently necessary to regard certain
familiar spacetimes as wild. For example, asymptotically flat spacetimes
diffeomorphic to ${\bf R}^{n}$ are usually treated as wild. The reason for
this is to keep Lorentz transformations distinct from gauge transformations,
by requiring the gauge parameters $\delta \xi^{\mu}$ to vanish at infinity.
Flat Minkowski spacetime is then not a boundary point of
Lor(${\bf R}^{n}$)/Diff(${\bf R}^{n}$) because the Poincar\'e isometries 
are not regarded as being contained in Diff(${\bf R}^{n}$).

\section{Metrics on configuration space}

It turns out to be both possible and useful to regard $\Phi$ and
${\Phi}/G$ not merely as differentiable (super)manifolds but as
pseudo-Riemannian (super)manifolds as well. Let $d\varphi^{i}$ be an
infinitesimal displacement in $\Phi$. We may associate with this
displacement a (super) arc lengths $ds$, given by
\begin{equation}
ds^{2}=d\varphi^{i} \; { }_{i}\gamma_{j} \; d\varphi^{j},
\label{(37)}
\end{equation}
where ${ }_{i}\gamma_{j}$ are the components of a (super) metric tensor
on $\Phi$. The ${ }_{i}\gamma_{j}$ are functionals of $\varphi$ having
the symmetry
$$
{ }_{i}\gamma_{j}=(-1)^{i+j+ij} { }_{j}\gamma_{i}
$$
and forming an invertible continuous matrix. The inverse, denoted by
$\gamma^{ij}$, satisfies
\begin{equation}
{ }_{i}\gamma_{k} \gamma^{kj}=\delta_{i}^{\; j}, \;
\gamma^{ik} \; { }_{k}\gamma_{j}=\delta_{\; j}^{i}, \;
\gamma^{ij}=(-1)^{ij}\gamma^{ji}.
\label{(38)}
\end{equation}

If the metric ${ }_{i}\gamma_{j}$ is chosen in such a way that the actions
of $G$ on $\Phi$ are isometries then ${ }_{i}\gamma_{j}$ induces also a
metric on the orbit space $\Phi / G$. One simply defines the distance
between neighbouring orbits in $\Phi / G$ to be the orthogonal distance 
between them in $\Phi$. This requires selecting ${ }_{i}\gamma_{j}$
in such a way that the continuous matrix $(-1)^{\alpha (i+1)}
Q_{\; \alpha}^{i} \; { }_{i}\gamma_{j} Q_{\; \beta}^{j}$ is nonsingular,
on all orbits so that a vector cannot be simultaneously tangent to and
orthogonal to any of them. 

Every element of $G$ infinitesimally close to the identity generates a
vector field $Q_{\; \alpha}^{i} \delta \xi^{\alpha}$ on $\Phi$ (see eq.
(21)). Each of these fields is a linear combination of the basic vector
fields $Q_{\; \alpha}^{i}$, and the structure of $G$ is determined by the
(super) Lie bracket relations (29) that they satisfy. The condition that
$G$ acts isometrically on $\Phi$ may be translated into the statement
that the (super) Lie derivatives of the metric ${ }_{i}\gamma_{j}$
with respect to the $Q's$ all vanish:
\begin{equation}
0={ }_{i}\gamma_{j}{\rvec {\cal L}}_{Q_{\alpha}}
={ }_{i}\gamma_{j,k} \; Q_{\; \alpha}^{k}
+(-1)^{\alpha(j+k)} { }_{i,}Q _{\; \alpha}^{k} \;
{ }_{k}\gamma_{j}+(-1)^{\alpha j}{ }_{i}\gamma_{k}
\; Q_{\; \alpha ,j}^{k}.
\label{(39)}
\end{equation}
It is not difficult to verify that eq. (29) is the integrability
condition for (39). Equation (39) generally has an infinity of solutions
differing nontrivially from one another. If it did not, i.e. if the 
solution were unique up to a constant factor, this would mean that $G$
acts transitively on $\Phi$ and hence that $\Phi / G$ is trivial, the
theory having no physical content.

In order to understand what eq. (39) says in more familiar terms it is
helpful to note that the fields $\varphi^{i}$ encountered in practice 
usually provide {\bf linear} realizations of their gauge groups. This is
cetainly true for the Yang-Mills and gravitational fields. What it means
is that the functional derivatives $Q_{\; \alpha,j}^{i}$ are independent
of the $\varphi^{i}$ and, when regarded as continuous matrices (in $i$ 
and $j$), yield a matrix {\bf representation} of the (graded) Lie algebra
associated with the group.

Of course this simplicity is generally lost if the $\varphi$'s are 
replaced by nonlinear functions of themselves. But it is remarkable that
there is usually a ``natural'' set of field variables of which the
$Q$'s are linear functionals. In gravity theory, in fact, there is a
{\bf family} of ``natural'' fields, namely all tensor densities of the
form
\begin{equation}
{\cal G}^{\mu \nu} \equiv g^{r} g^{\mu \nu} \; {\rm or} \; 
{\cal G}_{\mu \nu} \equiv g^{-r} g_{\mu \nu}, \; 
r \not = 1/n,
\label{(40)}
\end{equation}
where $g \equiv -{\rm det}(g_{\mu \nu})$.

Consider now the way in which the $Q$'s themselves change under 
infinitesimal gauge transformations. Using eq. (29) one finds
\begin{eqnarray}
\delta Q_{\; \alpha}^{i}&=& Q_{\; \alpha,j}^{i} \delta \varphi^{j}
=Q_{\; \alpha,j}^{i} Q_{\; \beta}^{j} \delta \xi^{\beta} \nonumber \\
&=& (-1)^{\alpha \beta}\Bigr(Q_{\; \beta,j}^{i} Q_{\; \alpha}^{j}
-Q_{\; \gamma}^{i} C_{\; \beta \alpha}^{\gamma}\Bigr)
\delta \xi^{\beta},
\label{(41)}
\end{eqnarray}
which says that $Q_{\; \alpha}^{i}$ is a two-point function that
transforms at the point associated with $i$ according to the representation
generated by the matrices $\Bigr(Q_{\; \alpha,j}^{i}\Bigr)$ and at the
point associated with $\alpha$ contragrediently to the representation 
generated by the matrices $\Bigr(C_{\; \gamma \beta}^{\alpha}\Bigr)$. In
gravity theory this says that the function $Q_{\mu \nu \sigma'}$ of eq.
(28) transforms like a covariant tensor at $x$ and like a covariant vector
density of unit weight at $x'$,\footnote{Under the diffeomorphism group
a tensor density of weight $w$, having $p$ covariant and $q$ contravariant
indices, transforms contragrediently to a tensor density of weight $1-w$,
having $q$ covariant and $p$ contravariant indices.} which indeed it does.
Equation (41) yields an analogous statement about the transformation law of
the two-point function $Q_{\; \mu \beta'}^{\alpha}$ of eq. (22) under the
Yang-Mills group. 

We are now ready to interpret eq. (39). Under the gauge group the
$\gamma$'s change according to
\begin{eqnarray}
\delta \; { }_{i}\gamma_{j}&=& { }_{i}\gamma_{j,k} \; \delta \varphi^{k}
={ }_{i}\gamma_{j,k} \; Q_{\; \alpha}^{k} \delta \xi^{\alpha} \nonumber \\
&=& -(-1)^{\alpha j}\Bigr[(-1)^{\alpha k}
{ }_{i,}Q_{\; \alpha}^{k} \; { }_{k}\gamma_{j}
+{ }_{i}\gamma_{k} \; Q_{\; \alpha,j}^{k}\Bigr]\delta \xi^{\alpha},
\label{(42)}
\end{eqnarray}
which says that the $\gamma$'s are two-point functions that transform
at each point contragrediently to the representation generated by the
matrices $\Bigr(Q_{\; \alpha,j}^{i}\Bigr)$. In gravity theory this
implies that when the $\varphi^{i}$ are chosen to be the components of the
covariant metric tensor, ${ }_{i}\gamma_{j}$ must transform at each point 
like a symmetric contravariant tensor density of unit weight. Any 
$\gamma$'s that transform in this way, and have an inverse $\gamma^{ij}$,
provide an acceptable metric on Lor($M$).

Among all such metrics on Lor($M$) there is a unique (up to a constant
factor) 1-parameter family of them that may be characterized as local.
These are given by
\begin{equation}
\gamma^{\mu \nu \sigma' \tau'} = \gamma^{\mu \nu \sigma \tau}\delta(x,x'),
\label{(43)}
\end{equation}
\begin{equation}
\gamma^{\mu \nu \sigma \tau} \equiv {1\over 2}g^{1/2}
\Bigr(g^{\mu \sigma}g^{\nu \tau}+g^{\mu \tau}g^{\nu \sigma}
+\lambda g^{\mu \nu}g^{\sigma \tau}\Bigr), \;
\lambda \not = -{2\over n}.
\label{(44)}
\end{equation}
For Yang-Mills theory in flat spacetime the corresponding metric is
\begin{equation}
\gamma_{\alpha \; \; \beta'}^{\; \mu \; \; \nu'}
=\gamma_{\alpha \beta}\eta^{\mu \nu}\delta(x,x'),
\label{(45)}
\end{equation}
where $\eta^{\mu \nu}$ is the Minkowski metric. The matrices
$(-1)^{\alpha (i+1)}Q_{\; \alpha}^{i} { }_{i}\gamma_{j}Q_{\; \beta}^{j}$
in the two cases are readily calculated to be respectively
\begin{equation}
-2g^{1/2}\left \{ \delta_{\mu \nu';\sigma}^{\; \; \; \; \; \; \; \; \sigma}
+R_{\mu}^{\; \sigma} \delta_{\sigma \nu'}
-(1+\lambda)[\delta(x,x')]_{; \mu \nu'} \right \}
\label{(46)}
\end{equation}
and 
\begin{equation}
-\delta_{\alpha \beta'; \mu}^{\; \; \; \; \; \; \; \; \mu}.
\label{(47)}
\end{equation}
The continuous matrix (47) is effectively the negative of the 
Yang-Mills-invariant Laplace-Beltrami operator. If the Yang-Mills field
is untwisted (see the lectures by Avis and Isham in this volume) it is
a nonsingular operator having a unique Green's function for each choice 
of boundary conditions at infinity. If the Yang-Mills field is twisted, 
however it may have zero eigenvalues, which means that the choice (45) 
fails to yield a globally valid metric on the orbit manifold. Although 
this is an important and interesting situation we shall not attempt to
deal with it in these lectures.

The continuous matrix (46) too may become singular. Its structure is
simplest when $\lambda=-1 (n \not = 2)$; it is then effectively a
slightly generalized form of the standard Laplace-Beltrami operator. 
Considerable evidence exists to indicate that it is nonsingular when the
spacetime manifold is diffeomorphic to ${\bf R}^{n}$. But for other
topologies it may have zero eigenvalues. Again we exclude this situation
from consideration.
 
When the matrices (46) and (47) are nonsingular, expressions (43) and
(45) constitute metrics on the space of fields which define, by orthogonal
projection, globally valid nonsingular metrics on the space of orbits. It
is possible to develop a theory of geodesics on these configuration spaces.
The geometry defined by (45) is flat and the geodesics in the space of
Yang-Mills potentials are trivial. The geometry defined by (43) and (44),
on the other hand, is not flat, and the resulting theory of geodesics on
the space of metric tensors $g_{\mu \nu}$ is not trivial. It can nevertheless
be shown that any pair of points in this space can be connected by a unique
geodesic. It can also be shown that if a geodesic intersects one orbit 
orthogonally then it intersects every orbit in its path orthogonally, and,
moreover, traces out a geodesic curve in the space of orbits. Methods for
proving these theorems can be found in DeWitt (1967a). Using these theorems
together with the fact that a vector in the space of metric tensors cannot
be simultaneously parallel and orthogonal to an orbit, one can then prove
that any pair of orbits can be connected by a unique geodesic. It should
be stated that all of these theorems depend upon the maintenance of fixed
boundary conditions (on fields and diffeomorphisms) at infinity.

\section{Volume elements on configuration space}

With a metric defined on the space of fields it is possible to introduce
a formal volume element $\mu d\varphi$ ($d\varphi \equiv \prod_{i}
d\varphi^{i}$) by choosing
\begin{equation}
\mu={\rm const}. \times |{\rm det}({ }_{i}\gamma_{j})|^{1/2}.
\label{(48)}
\end{equation}
This volume element is gauge invariant and can be used to define gauge
invariant functional integrals over configuration space. When fermionic
fields are present the determinant in eq. (48) is the {\bf super}
determinant (see Nath (1976)), which satisfies the variational law  
\begin{equation}
\delta {\rm ln} \; {\rm det} ({ }_{i}\gamma_{j})=(-1)^{i}\gamma^{ij}
\delta \; { }_{j}\gamma_{i}.
\label{(49)}
\end{equation}
This law, combined with eq. (39), yields the following equation of
``divergenceless flow'' that could in principle be used to select a 
gauge invariant volume element independently of a metric:
\begin{equation}
(-1)^{i(\alpha+1)}\Bigr(\mu Q_{\; \alpha}^{i}\Bigr)_{,i}=0.
\label{(50)}
\end{equation}

The delta functions contained in the metrics (43) and (45) give these
metrics a block structure that yields simple formal expressions for their
determinants. In Yang-Mills theory the determinant is a constant; in 
gravity theory it is given by
\begin{equation}
{\rm det}\Bigr(\gamma^{\mu \nu \sigma' \tau'}\Bigr)=\prod_{x}\gamma(x),
\label{(51)}
\end{equation}
where $\gamma(x)$ is the determinant of the ${1\over 2}n(n+1) \times
{1\over 2}n(n+1)$ matrix $\gamma^{\mu \nu \sigma \tau}$. It is not a
difficult computation to show that
\begin{equation}
\gamma=(-1)^{n-1}\left(1+{n \lambda \over 2}\right)
g^{{1\over 4}(n-4)(n+1)}.
\label{(52)}
\end{equation}
In a 4-dimensional spacetime $\gamma$, and hence 
${\rm det}(\gamma^{\mu \nu \sigma' \tau'})$, is seen to be a constant,
independent of the $g_{\mu \nu}$. The functional $\mu$, in the volume
element over the configuration space of gravity theory, may therefore
be taken to be a constant. Without loss of generality it may be chosen
equal to $1$. This will no longer be true in other dimensions, or when
other fields are present in addition to the gravitational field, if we
stick to the $g_{\mu \nu}$ as the basic field variables. However, we can
in principle replace the $g_{\mu \nu}$ by one of the family of variables
defined in eq. (40) and choose $r$ so that $\mu$ remains constant. In
practice, as we shall see later, this is unnecessary. To set $\mu$ 
``effectively'' equal to unity it turns out to be necessary only to choose
basic fields that transform linearly under the gauge group.

\section{Group coordinates}

The scalar functions $\xi^{\alpha}(x)$ that parameterize the elements
of the Yang-Mills group may be regarded as ``coordinates'' in the group
manifold. In the case of the diffeomorphism group the group coordinates
may be taken to be the functions $\xi^{\mu}(x)$ that define the coordinate
transformation $x^{\mu} \rightarrow \xi^{\mu}$ associated with each 
diffeomorphism in each coordinate chart or patch. Note that the functions
$\xi^{\mu}(x)$ are neither scalars nor components of vectors. Note also 
that in both cases the coordinatization of the group cannot generally be
achieved without bringing in the whole apparatus of charts, atlases and
consistency conditions in the regions of intersection of overlapping charts.

Any group may be regarded as acting on itself through multiplication either
on the left or on the right. Every group thus provides a realization of
itself, and if it is a gauge group $G$, possesses a set of functionals
$Q_{\; \beta}^{\alpha}[\xi]$ analogous to the functionals
$Q_{\; \alpha}^{i}[\phi]$ over the configuration space $\Phi$. The
$Q_{\; \beta}^{\alpha}$ are then defined by
\begin{equation}
[(I+\delta \xi)\xi]^{\alpha}=\xi^{\alpha}+Q_{\; \beta}^{\alpha}[\xi]
\delta \xi^{\beta} \; {\rm for} \; {\rm all} \; \delta \xi^{\alpha}
\; {\rm and} \; {\rm all} \; \xi \; {\rm in} \; G,
\label{(53)}
\end{equation}
where $I$ denotes the identity element of $G$ and $I+\delta \xi$ denotes
an element of $G$ whose coordinates differ by infinitesimal amounts 
$\delta \xi^{\alpha}$ from those of $I$. Using the fact that 
$I^{\mu}(x)=x^{\mu}$, and that $(\xi \xi')^{\mu}(x)=\xi^{\mu}(\xi(x))$
for all $\xi$ and $\xi'$ in $G$, it is not difficult to verify that the
$Q_{\; \beta}^{\alpha}$ for the diffeomorphism group are given
explicitly by
\begin{equation}
Q_{\; \nu'}^{\mu}[\xi]=\delta_{\; \nu}^{\mu} \; \delta(\xi(x),x').
\label{(54)}
\end{equation}
In the case of the Yang-Mills groups the $Q_{\; \beta}^{\alpha}$ take
the forms
\begin{equation}
Q_{\; \beta'}^{\alpha}[\xi]=g_{\; \beta}^{\alpha}(\xi(x))\delta(x,x'),
\label{(55)}
\end{equation}
where the $g_{\; \beta}^{\alpha}$ are the corresponding quantities for 
the associated Lie group.

Now let $\delta \xi^{\alpha}=\xi^{\alpha} \delta t$ for some fixed 
$\xi^{\alpha}$ and consider the curve in $G$ defined by
\begin{equation}
\xi(t)=\lim_{\delta t \to 0}(I+\delta \xi)^{t / \delta t}.
\label{(56)}
\end{equation}
Evidently 
\begin{equation}
\xi(s)\xi(t)=\xi(t)\xi(s)=\xi(s+t), \;
\xi(0)=I, \; \xi^{-1}(t)=\xi(-t),
\label{(57)}
\end{equation}
\begin{equation}
{d \xi^{\alpha}(t)\over dt}=Q_{\; \beta}^{\alpha}[\xi(t)]\xi^{\beta}.
\label{(58)}
\end{equation}
The points on the curve are seen to constitute a one-parameter Abelian
subgroup of $G$.

If it could be proved that all the elements of $G$ in a neighbourhood 
$N$ of the identity can be obtained by a process of exponentiation of
the form (56) then it would follow that the one-parameter Abelian 
subgroups completely span the neighbourhood $N$. A special set of 
coordinates $\xi_{c}^{\; \alpha}$, known as {\bf canonical coordinates},
could be introduced in $N$ for which the functions $\xi^{\alpha}(t)$ above
take the simple form 
\begin{equation}
\xi_{c}^{\; \alpha}(t)=\xi^{\alpha}t.
\label{(59)}
\end{equation}
Let us assume that our coordinates are already canonical, so that we
may drop the subscript $c$. Then we have
\begin{equation}
I^{\alpha}=0, \; {\xi^{-1}}^{\alpha}=-\xi^{\alpha},
\label{(60)}
\end{equation}
\begin{equation}
\xi^{\alpha}=Q_{\; \beta}^{\alpha}[\xi]\xi^{\beta}
=Q_{\; \; \; \; \; \beta}^{-1 \alpha}[\xi]\xi^{\beta},
\label{(61)}
\end{equation}
where $\Bigr(Q_{\; \; \; \; \; \beta}^{-1 \alpha}\Bigr)$ is the (continuous)
matrix inverse to $\Bigr(Q_{\; \beta}^{\alpha}\Bigr)$. By taking note
of the fact that the $Q_{\; \beta}^{\alpha}$ must satisfy an identity
analogous to (29), namely
\begin{equation}
Q_{\; \alpha, \delta}^{\gamma} \; Q_{\; \beta}^{\delta}
-(-1)^{\alpha \beta}Q_{\; \beta,\delta}^{\gamma} \;
Q_{\; \alpha}^{\delta}=Q_{\; \delta}^{\gamma} \; 
C_{\; \alpha \beta}^{\delta},
\label{(62)}
\end{equation}
we may show that in a canonical coordinate system the $Q_{\; \beta}^{\alpha}$
are completely determined by the structure constants.

We begin by rewriting eq. (62) in the equivalent form
\begin{equation}
Q_{\; \; \; \; \; \beta,\gamma}^{-1 \alpha}-(-1)^{\beta \gamma}
Q_{\; \; \; \; \; \gamma,\beta}^{-1 \alpha}
+(-1)^{\epsilon (\delta+\beta)} C_{\; \delta \epsilon}^{\alpha}
Q_{\; \; \; \; \; \beta}^{-1 \delta} 
\; Q_{\; \; \; \; \; \gamma}^{-1 \epsilon}=0.
\label{(63)}
\end{equation}
Multiplying this equation on the right by $\xi^{\alpha}$ and using eq.
(61) we get
\begin{equation}
Q_{\; \; \; \; \; \beta,\gamma}^{-1 \alpha}\xi^{\gamma}
-(-1)^{\beta \gamma}Q_{\; \; \; \; \; \gamma,\beta}^{-1 \alpha}\xi^{\gamma}
+C_{\; \delta \epsilon}^{\alpha} \xi^{\epsilon} 
Q_{\; \; \; \; \; \beta}^{-1 \delta}=0.
\label{(64)}
\end{equation}
On the other hand, differentiating eq. (61) with respect to $\xi^{\beta}$
we find
\begin{equation}
(-1)^{\beta \gamma}Q_{\; \; \; \; \; \gamma,\beta}^{-1 \alpha} \xi^{\gamma}
+Q_{\; \; \; \; \; \beta}^{-1 \alpha}=\delta_{\; \beta}^{\alpha}.
\label{(65)}
\end{equation}
Addition of eqs. (64) and (65) yields
\begin{equation}
Q_{\; \; \; ,\alpha}^{-1}\xi^{\alpha}+Q^{-1}- C \cdot \xi Q^{-1}=1,
\label{(66)}
\end{equation}
where ``1'' denotes the unit matrix (delta function) and
\begin{equation}
Q^{-1}[\xi] \equiv \left(Q_{\; \; \; \; \; \beta}^{-1 \alpha}[\xi]\right), \;
C \cdot \xi \equiv (-1)^{\beta \gamma}C_{\; \gamma \beta}^{\alpha}
\xi^{\gamma}=-\Bigr(C_{\; \beta \gamma}^{\alpha} \xi^{\gamma}\Bigr).
\label{(67)}
\end{equation}
The solution of eq. (66) satisfying the necessary boundary condition
\begin{equation}
Q_{\; \beta}^{\alpha}[I]=\delta_{\; \beta}^{\alpha}
\label{(68)}
\end{equation}
(see eq. (53)) is
\begin{equation}
Q^{-1}[\xi]={e^{C \cdot \xi}-1 \over C \cdot \xi} \equiv
1+{1\over 2!}C \cdot \xi +{1\over 3!}(C \cdot \xi)^{2}+... \; .
\label{(69)}
\end{equation}

The series (69) converges for all values of the $\xi^{\alpha}$. For certain
values the (continuous) matrix $Q^{-1}$ may have vanishing roots. For these
values some of the $Q_{\; \beta}^{\alpha}$, and hence the canonical 
coordinate system itself, become singular. In the case of an untwisted
Yang-Mills group it can be shown that the one-parameter Abelian subgroups
do span a neighbourhood of the identity. (This is, in fact, a corollary
of the corresponding theorem for the associated Lie group.) Indeed they
span the entire group - or, rather, that part of the group that is connected
to the identity, i.e. the proper group. The whole group can therefore be
parameterized by canonical coordinates (supplemented, perhaps, with some
discrete labels).

Canonical coordinates for the Yang-Mills group have a periodic, or angular,
nature. At a given point $x$ of the spacetime manifold let the 
$\xi^{\alpha}(x)$ in eq. (55) increase in magnitude but maintain fixed
ratios to one another. Eventually all of the $g_{\; \beta}^{\alpha}$ will
become singular at once. One has returned to the identity element of the
associated Lie group. By allowing the canonical coordinates to range
from $-\infty$ to $\infty$ one evidently covers the gauge group an infinity
of times. Despite the fact that the $Q_{\; \beta}^{\alpha}$ become 
singular for certain values of the $\xi's$, the canonical coordinates
are good in that, no matter what their values, they always define a unique
element of the group.

\section{No canonical coordinates for the diffeomorphism group}

If canonical coordinates could be introduced into the diffeomorphism 
group one could dispense with the apparatus of charts, atlases, etc. in
parameterizing the group. Every diffeomorphism could be characterized by
a (finite) vector field just as those infinitesimally close to the identity
can be characterized by an infinitesimal vector field. And a vector field
has a meaning independent of charts and atlases.

Unfortunately the one-parameter Abelian subgroups of the diffeomorphism
group do not span a neighbourhood of the identity. If the dimensionality
$n$ of the spacetime manifold $M$ is greater than or equal to $2$ there
are $C^{\infty}$ diffeomorphisms arbitrarily close to the identity that
cannot be obtained by exponentiation as in eq. (56). The proof, which we
now outline, was first given by Freifeld (1968).

It suffices to confine attention to ${\bf R}^{2}$ or, equivalently,
to the complex plane ${\bf C}$. Let $x$ be a point of ${\bf C}$. Instead
of breaking $x$ into its real and imaginary parts we may treat $x$ and
its complex conjugate $x^{*}$ formally as independent variables. A
$C^{\infty}$ diffeomorphism $\xi: {\bf C} \rightarrow {\bf C}$ is then
a one-to-one complex function $\xi(x,x^{*})$, of class $C^{\infty}$ in
both $x$ and $x^{*}$, whose inverse, $x(\xi,\xi^{*})$ is $C^{\infty}$
in $\xi$ and $\xi^{*}$.

Let $N$ be a positive integer and $\alpha$ a positive real number. 
Suppose $\xi$ has the analytic form 
\begin{equation}
\xi(x,x^{*})=e^{2 \pi i \over N} x + \alpha x^{N+1}
\label{(70)}
\end{equation}
in a finite neighbourhood of the origin (e.g., in a circle of finite
radius), and suppose that outside of this neighbourhood $\xi$ changes
smoothly ($C^{\infty}$) to the identity function $\xi(x,x^{*})=x$. If
$N$ is chosen large and $\alpha$ is chosen small then $\xi$ and all its
derivatives may be made uniformly close to those of the identity. We
shall show that $\xi$ does not lie on a one-parameter subgroup of
$C^{\infty}$ diffeomorphisms $\xi(t):{\bf C} \rightarrow {\bf C}$
with $\xi(0)=I$.

Suppose we assume that it does lie on such a subgroup. Without loss of
generality we may also assume that $\xi(1)=\xi$, and then we have
\begin{equation}
\xi(0,x,x^{*})=x, \; \xi(1,x,x^{*})=\xi(x,x^{*}),
\label{(71)}
\end{equation}
as well as
\begin{equation}
\xi(s,\xi(t,x,x^{*}),\xi^{*}(t,x,x^{*}))
=\xi(t,\xi(s,x,x^{*}),\xi^{*}(s,x,x^{*}))
=\xi(s+t,x,x^{*}).
\label{(72)}
\end{equation}
Note that the diffeomorphism (70) leaves the origin fixed. Therefore
\begin{equation}
\xi(0,0,0)=0, \; \xi(1,0,0)=0.
\label{(73)}
\end{equation}
Define
\begin{equation}
z(t) \equiv \xi(t,0,0).
\label{(74)}
\end{equation}
The function $z(t)$ describes a closed curve passing through the origin
in the complex plane. Using eqs. (72) and (73) we find
\begin{eqnarray}
\; & \; & \xi(z(t),z^{*}(t))=\xi(1,z(t),z^{*}(t))
=\xi(1,\xi(t,0,0),\xi^{*}(t,0,0)) \nonumber \\
&=& \xi(t,\xi(1,0,0),\xi^{*}(1,0,0))
=\xi(t,0,0)=z(t),
\label{(75)}
\end{eqnarray}
which implies that the diffeomorphism (70) leaves every point on this
closed curve fixed. But the only curve passing through the origin that
(70) leaves fixed is the degenerate curve consisting of the single point 
$x=0$. Therefore every one of the diffeomorphisms $\xi(t)$ must leave
the origin fixed:
\begin{equation}
\xi(t,0,0)=0 \; {\rm for} \; {\rm all} \; t.
\label{(76)}
\end{equation}

Since $\xi$ and the $\xi(t)$ are $C^{\infty}$ we may consider their 
formal Taylor series at the origin. The formal Taylor series for
$\xi$, which is just expression (70), must lie on the one-parameter
group of formal Taylor series for the $\xi(t)$, which may be written 
in the form
\begin{equation}
\xi(t,x,x^{*})=\sum_{m,n=0}^{\infty}a_{m,n}(t)x^{m}x^{* \; n}.
\label{(77)}
\end{equation}
Furthermore, these formal Taylor series must satisfy (formally) eqs. 
(72).

In view of eqs. (71) and (76) it is evident that
\begin{eqnarray}
\; & \; & a_{0,0}(t)=0 \; {\rm for} \; {\rm all} \; t; \nonumber \\
& \; & a_{1,0}(0)=1, \; {\rm all} \; {\rm other} \; a_{m,n}(0)'s \;
{\rm vanish}; \nonumber \\
& \; & a_{1,0}(1)=e^{2 \pi i \over N}, \;
a_{N+1,0}(1)=\alpha, \; {\rm all} \; {\rm other} \;
a_{m,n}(1)'s \; {\rm vanish}.
\label{(78)}
\end{eqnarray}
Moreover, inserting (77) into (72) with $s=t={1\over 2}$, one finds
$$
e^{2 \pi i \over N}x + \alpha x^{N+1}
=\sum_{m,n=0}^{\infty}a_{m,n}\left({1\over 2}\right)
\left[\xi \left({1\over 2},x,x^{*}\right)\right]^{m}
\left[\xi^{*}\left({1\over 2},x,x^{*}\right)\right]^{n} 
$$
$$ 
= a_{1,0}\left({1\over 2}\right)\left[a_{1,0}\left({1\over 2}\right)x
+a_{0,1}\left({1\over 2}\right)x^{*}+... \right]
+ a_{0,1}\left({1\over 2}\right)\left[
a_{1,0}^{*}\left({1\over 2}\right)x^{*}
+a_{0,1}^{*}\left({1\over 2}\right)x+... \right]+...,
$$
whence
\begin{equation}
e^{2\pi i \over N}=\left[a_{1,0}\left({1\over 2}\right)\right]^{2}
+|a_{0,1}(1/2)|^{2},
\label{(79)}
\end{equation}
\begin{equation}
0=a_{0,1}\left({1\over 2}\right) 
{\rm Re} a_{1,0}\left({1\over 2}\right).
\label{(80)}
\end{equation}
Suppose $a_{0,1}\left({1\over 2}\right) \not = 0$. Then 
$a_{1,0}\left({1\over 2}\right)$ must be pure imaginary, and the right hand
side of eq. (79) must be a real number, which contradicts the left hand
side. Therefore
$$
a_{0,1}\left({1\over 2}\right)=0, \; 
a_{1,0}\left({1\over 2}\right)
=e^{{1\over 2}\left({2\pi i \over N}+2\pi i K \right)},
$$  
for some integer $K$. Repeating this reasoning for $s=t={1\over 4},
s=t={1\over 8}$, etc., one obtains, by continuity, 
\begin{equation}
a_{0,1}(t)=0 \; {\rm for} \; {\rm all} \; t, \; a_{1,0}(t)=e^{\beta t}, \;
\beta=2\pi i \left({1\over N}+K \right).
\label{(81)}
\end{equation}

We now have
\begin{equation}
\xi(t,x,x^{*})=e^{\beta t}x+\sum_{m+n \geq 2} a_{m,n}(t)x^{m}x^{* \; n}.
\label{(82)}
\end{equation}
Insertion of this formal series into (72) yields
\begin{equation}
a_{m,n}(s+t)=e^{\beta s}a_{m,n}(t)+e^{(m-n)\beta t}a_{m,n}(s), \;
m+n=2.
\label{(83)}
\end{equation}
This functional equation can be solved by differentiating with respect to 
$s$ and setting $s=0$:
\begin{equation}
\left({d\over dt}-\beta \right)a_{m,n}(t)={\dot a}_{m,n}(0)
e^{(m-n)\beta t}, \; m+n=2.
\label{(84)}
\end{equation}
(Here the dot denotes the derivative.) The Green's function for the
operator ${d\over dt}-\beta$ appropriate to the boundary conditions (78)
is
$$
\Bigr[\theta(t-t')\theta(t')-\theta(t'-t)\theta(-t')\Bigr]
e^{\beta (t-t')}
$$
where $\theta$ is the step function. Use of this Green's function yields
\begin{equation}
a_{m,n}(t)={{\dot a}_{m,n}(0)\over (m-n-1)\beta}
\Bigr[e^{(m-n)\beta t}-e^{\beta t}\Bigr], \; m+n=2,
\label{(85)}
\end{equation}
which is easily verified to satisfy (83). Now if $N$ is large 
$a_{m,n}(1)$ ($m+n=2$) must vanish in virtue of the last of eqs. (78).
But the right hand side of (85) does not vanish at $t=1$ unless
${\dot a}_{m,n}(0)=0$. Therefore
\begin{equation}
a_{m,n}(t)=0 \; {\rm for} \; {\rm all} \; t \; {\rm when} \;
m+n=2,
\label{(86)}
\end{equation}
and hence
\begin{equation}
\xi(t,x,x^{*})=e^{\beta t}x+\sum_{m+n \geq 3}a_{m,n}(t)x^{m}x^{* \; n}.
\label{(87)}
\end{equation}

Inserting {\bf this} series into (72) one gets
\begin{equation}
a_{m,n}(s+t)=e^{\beta s}a_{m,n}(t)+e^{(m-n)\beta t}a_{m,n}(s), \;
m+n=3,
\label{(88)}
\end{equation}
which is identical with eq. (83) except that now $m+n=3$. The solution
is the same as before:
\begin{equation}
a_{m,n}(t)={{\dot a}_{m,n}(0)\over (m-n-1)\beta}
\Bigr[e^{(m-n)\beta t}-e^{\beta t}\Bigr], \; m+n=3.
\label{(89)}
\end{equation}
It is now possible for the factor $m-n-1$ in the denominator to vanish,
in which case this solution is replaced by its limit as
$m-n \rightarrow 1$:
\begin{equation}
a_{m,n}(t)={\dot a}_{m,n}(0)t e^{\beta t}, \; m-n=1.
\label{(90)}
\end{equation}
Once again, comparing (89) and (90) with the boundary condition 
$a_{m,n}(1)=0$ ($m+n=3$), one must conclude that
\begin{equation}
a_{m,n}(t)=0 \; {\rm for} \; {\rm all} \; t \; {\rm when} \; m+n=3.
\label{(91)}
\end{equation}
In fact, continuing in this way one finds 
\begin{equation}
a_{m,n}(t)=0 \; {\rm for} \; {\rm all} \; t \; {\rm and} \; {\rm all}
\; m,n \; {\rm with} \; 2 \leq m+n \leq N. 
\label{(92)}
\end{equation}

One arrives finally at the case $m+n=N+1$, where one obtains
\begin{equation}
a_{N+1,0}(t)={1\over N \beta}{\dot a}_{N+1,0}(0)e^{\beta t}
\Bigr(e^{N \beta t}-1 \Bigr).
\label{(93)}
\end{equation}
This expression {\bf vanishes} at $t=1$, precisely where we don't want
it to! According to (78) we must have $a_{N+1,0}(1)=\alpha$. We have
thus arrived at a contradiction. Q.E.D.

Since only a small (infinitesimal) neighbourhood of the origin is really
involved in the above analysis, it follows that the restriction to
${\bf R}^{2}$ is not essential. For any differentiable manifold of 
dimension greater than or equal to $2$ there exist $C^{\infty}$ 
diffeomorphisms arbitrarily close to the identity that do not lie on
one-parameter subgroups of $C^{\infty}$ diffeomorphisms.

\section{Gauge conditions}

A {\bf gauge condition} is a set of constraints that picks out a 
subspace in the configuration space $\Phi$, of codimension equal to the
dimension of the gauge group $G$. The gauge condition is said to be
{\bf globally valid} if this subspace intersects each orbit in precisely
one point. Such a subspace exists in the Yang-Mills case only if the
gauge group corresponds to an untwisted fiber bundle. For the 
diffeomorphism group it probably exists if spacetime is diffeomorphic
to ${\bf R}^{n}$. We confine our attention to these cases. The 
subspace may then be regarded as {\bf representing} the orbit manifold
$\Phi / G$. Each orbit is represented by the point at which it intersects
the subspace.

To express this idea in equations one may think of the variables 
$\varphi^{i}$ as being replaced by other variables $I^{A},P^{\alpha}$,
where the $I^{A}$ label individual orbits and are gauge invariant, and
the $P^{\alpha}$ label corresponding points {\bf in} each orbit. The point
on each orbit that is selected by the given gauge condition may be chosen
as the origin of the ``coordinates'' $P^{\alpha}$ in that orbit. The gauge
condition is then simply $P^{\alpha}=0$.

It will actually prove convenient to work with the continuum of gauge
conditions
\begin{equation}
P^{\alpha}[\varphi]=\zeta^{\alpha},
\label{(94)}
\end{equation}
where the $\zeta^{\alpha}$ are constants (i.e., independent of the
$\varphi^{i}$) whose values range over some preselected domain. Explicit
functional forms for the $P^{\alpha}$ in terms of the $\varphi^{i}$ may
be obtained (in principle) as follows. Remembering that each (generic)
orbit is a copy of $G$, choose the $P^{\alpha}$ to be a set of group
coordinates. Since the action of the gauge group on each (generic) orbit
mimics its action on itself such $P$'s must be solutions of the functional
differential equations\footnote{These equations are readily verified to be
integrable in virtue of the identities (29) and (62).}
\begin{equation}
P_{\; ,i}^{\alpha}[\varphi]Q_{\; \beta}^{i}[\varphi]
=Q_{\; \beta}^{\alpha}[P[\varphi]].
\label{(95)}
\end{equation}
The domain over which the $\zeta^{\alpha}$ in eq. (94) range may then be
taken to be the full domain of the group coordinates.

Equations (95) do not suffice completely to determine the $P^{\alpha}$.
Additional conditions are needed to ``line up'' corresponding points on
adjacent orbits. One possible way to do the lining up is as follows. 
Introduce into the configuration space $\Phi$ one of the metrics
${ }_{i}\gamma_{j}$ previously discussed. Choose a generic orbit and call
it the {\bf base orbit}. Call the identity element on that orbit the
{\bf base point}. Let $V$ be the subspace of $\Phi$ generated by the set
of all geodesics emanating from the base point in directions orthogonal
to the base orbit. As previously noted, these geodesics intersect all
orbits in their paths orthogonally. Using the fact that every pair of
points in $\Phi$ can be connected by a unique geodesic (at least in the
Yang-Mills and gravitational cases) and the fact that a geodesic cannot
be simultaneously orthogonal to and tangent to an orbit, one can show 
that $V$ ultimately intersects all orbits. To keep it from intersecting a 
given orbit more than once one may terminate each of the generating
geodesics as soon as it strikes a boundary point of $\Phi /G$. $V$ is
then topologically (but not necessarily metrically) a copy of $\Phi / G$.

To gain an appreciation of some of the metrical situations that can arise
think of $\Phi$ as being ${\bf R}^{3}$ and $G$ as being the group of screw
motions with fixed nonvanishing pitch about some axis. The orbits are then
helices and all, including the axis itself, are generic. If ${\bf R}^{3}$
bears the Cartesian metric then the orbit space $\Phi /G$ is
topologically but not metrically a plane. Note that in this example there
exist no surfaces that intersect all orbits orthogonally, although every
plane not containing the axis is perpendicular to some orbit at its
intersection point and is a surface like $V$, based on that orbit.

Returning now to the general problem, we may place the identity element on
each orbit at the point where the orbit intersects $V$. If another subspace
$V'$ is constructed like $V$ but starting from another point on the base
orbit, it too will intersect all the orbits. Because the group operations
are isometries of ${ }_{i}\gamma_{j}$, the $P^{\alpha}$ will be constants
over $V'$. That is, once the identity points are ``lined up'' all the other
points are automatically lined up too. The gauge condition (94) is 
therefore globally valid for all $\zeta^{\alpha}$ in the domain $G$.

Unfortunately in practice it is almost hopelessly difficult to implement
constructions like this one, which are guaranteed to yield globally valid
gauge conditions. In the present construction, because $V$ is generally
orthogonal to none but the base orbit, one is faced with the problem of
solving global functional constraints rather than functional differential
equations. Even the functional differential equations that one has, namely
Eqs. (95), are highly nontrivial. In the Yang-Mills and gravitational 
cases they take respectively the forms
\begin{equation}
\Bigr[\delta P^{\alpha}(x)/ \delta A_{\; \mu}^{\beta}(x')\Bigr]_{;\mu'}
={\cal G}_{\; \beta}^{\alpha}(P(x))\delta(x,x'),
\label{(96)}
\end{equation}
\begin{equation}
2 \Bigr[\delta P^{\mu}(x)/\delta g_{\nu \sigma}(x')\Bigr]_{;\sigma'}
=\delta_{\; \nu}^{\mu} \delta(P(x),x'),
\label{(97)}
\end{equation}
(see eqs. (22), (28), (54) and (55)).

By far the bulk of all work on Yang-Mills theory and quantum gravity has
made use of {\bf linear} gauge conditions, i.e., conditions (94) with
$P^{\alpha}$ taken in the form
\begin{equation}
P^{\alpha}[\varphi]=P_{\; i}^{\alpha}[\varphi_{B}]\phi^{i}, \;
\phi^{i}=\varphi^{i}-\varphi_{B}^{i}
\label{(98)}
\end{equation}
where $\varphi_{B}^{i}$ is some fiducial field, often called a
{\bf background field}. To ensure that the subspaces defined by (94) do
indeed intersect the orbits uniquely, at least in the vicinity of the
background field and with the $\zeta^{\alpha}$ close to zero, one often makes
use of the orthogonality idea by choosing 
\begin{equation}
P_{\; i}^{\alpha}[\varphi_{B}]=(-1)^{\alpha (j+1)}
Q_{\; \alpha}^{i}[\varphi_{B}] \; { }_{j}{\gamma}_{i}[\varphi_{B}].
\label{(99)}
\end{equation}
For example, if ${ }_{i}\gamma_{j}$ has the form (43) then, with the choice
(99), the condition $P^{\alpha}=0$ becomes
\begin{equation}
g_{B}^{1/2}\Bigr(2 \phi_{\mu}^{\; \nu}+\lambda \delta_{\mu}^{\; \nu}
\phi_{\sigma}^{\; \sigma}\Bigr)_{;\nu}=0, \;
\phi_{\mu \nu}=g_{\mu \nu}-g_{B \mu \nu}.
\label{(100)}
\end{equation}
Here indices are raised and lowered by means of the background metric
$g_{B \mu \nu}$ and the covariant derivative is defined in terms of it.
With $\lambda$ set equal to $-1$ (see the comments following eqs. (46)
and (47)) this is a very popular gauge condition in quantum gravity. The
corresponding condition in Yang-Mills theory, with
${ }_{i}\gamma_{j}$ given by (45), is
\begin{equation}
\phi_{\alpha \; \; ; \mu}^{\; \mu}=0, 
\label{(101)}
\end{equation}
\begin{equation}
\phi_{\; \mu}^{\alpha}=A_{\; \mu}^{\alpha}
-A_{B \; \; \mu}^{\; \; \alpha}.
\label{(102)}
\end{equation}
Here indices are raised and lowered by means of the metrics 
$\gamma_{\alpha \beta}$ and $\eta_{\mu \nu}$ and the covariant derivative
is defined in terms of the background field $A_{B \; \; \mu}^{\; \; \alpha}$.
Condition (101) is known as the {\bf Lorenz condition}.

Linear gauge conditions are extremely convenient in perturbation theory,
where the field $\varphi^{i}$ is treated as if it never gets very far from
the background $\varphi_{B}^{\; i}$. Covariant (with respect to the
background) gauge conditions like (100) and (101) are usually the best,
but for some purposes noncovariant gauges (e.g., the Coulomb gauge in
Yang-Mills theory) are more useful. In non-perturbative studies, however,
linear gauge conditions have to be used with great care (see Gribov (1977)).
At least five things can go wrong with linear gauge conditions when
applied globally: 
\vskip 0.3cm
\noindent
(1) The subspace defined by a linear condition may or
may not have a boundary, and if it does this boundary may not coincide
with the boundary (if any) of $\Phi /G$. 
\vskip 0.3cm
\noindent
(2) The subspace defined by a
linear condition may intersect some orbits more than once. 
\vskip 0.3cm
\noindent
(3) There may be some orbits that it does not intersect at all. 
\vskip 0.3cm
\noindent
(4) Even if it intersects all orbits when the $\zeta^{\alpha}$ in eq. (94)
have certain values, it may not intersect all orbits when the 
$\zeta^{\alpha}$ have other values. This means that there is no natural
domain for the $\zeta^{\alpha}$. 
\vskip 0.3cm
\noindent
(5) When $G$ is ``twisted'' there are no globally valid gauge conditions
at all, linear or otherwise.

If any of the above situations hold, the subspace defined by (94) will not
represent $\Phi / G$ faithfully. It is possible in some cases to patch
things up so that the advantages of linear gauge conditions can be 
maintained. This has been done in certain global studies in Yang-Mills
theory. However, the diffeomorphism group, as we have repeatedly 
emphasized, is a much more complicated group than the Yang-Mills group 
and both the difficulties to which it gives rise globally and the
opportunities that it presents for technical innovation are almost unknown
at the present time. In order to keep all options open we shall first
develop the formal theory using foolproof gauge conditions, such as those
based on group coordinates, and then make some remarks about how things
might go when other gauge conditions are used.

\section{The action. Vertex functions. Renormalizability.}

The dynamical behavior of any field is determined by its action functional
$S$. The action functionals of (pure) Yang-Mills and gravity theories 
are respectively
\begin{equation}
S_{A}=-{1\over 4}\int F_{\alpha \mu \nu}F^{\alpha \mu \nu}d^{n}x,
\label{(103)}
\end{equation}
\begin{equation}
S_{g}=2 \int g^{1/2}R d^{n}x.
\label{(104)}
\end{equation}
As long as the limits of integration are not specified these integrals
must be regarded as purely formal expressions that serve merely to
yield the dynamical equations:
\begin{equation}
0=\delta S_{A}/ \delta A_{\; \mu}^{\alpha} \equiv
-F_{\alpha \; \; \; ;\nu}^{\; \mu \nu},
\label{(105)}
\end{equation}
\begin{equation}
0=\delta S_{g}/ \delta g_{\mu \nu} \equiv
-2 g^{1/2}\left(R^{\mu \nu}-{1\over 2}g^{\mu \nu}R \right).
\label{(106)}
\end{equation}
For some purposes, however, values need to be assigned to the actions.
Integration boundaries must then be specified and, in the case of the
gravitational field, a surface integral must be split off from (104)
so that the integrand involves derivatives of $g_{\mu \nu}$ of order
no higher than the first.

We refer the student to standard references (e.g., Misner, Thorne and
Wheeler (1973)) for analyses of the initial value problems associated
with eqs. (105) and (106). From these analyses it is readily deduced
that in a spacetime of $n$ dimensions the Yang-Mills field has
$n-2$ degrees of freedom per spatial point and the gravitational
field has ${1\over 2}n(n-3)$.

In the generic notation, eqs. (103) and (104) are written
\begin{equation}
S_{,i}=0.
\label{(107)}
\end{equation}
Gauge invariance of the theory is guaranteed by the identity
\begin{equation}
S_{,i} Q_{\; \alpha}^{i} = 0,
\label{(108)}
\end{equation}
or, more explicitly,
\begin{equation}
F_{\alpha \; \; \; ; \nu \mu}^{\; \mu \nu}=0, \;
4 \left[g^{1/2}\left(R^{\mu \nu}-{1\over 2}g^{\mu \nu}R \right)
\right]_{; \nu} =0.
\label{(109)}
\end{equation}
The left hand side of eq. (107) transforms linearly under the gauge
group and hence the gauge group leaves the field equations intact. This
is most easily seen by functionally differentiating eq. (108), which
yields
\begin{equation}
\delta S_{,i}=S_{,ij}\delta \varphi^{j}
=S_{,ij}Q_{\; \alpha}^{j} \delta \xi^{\alpha}
=-(-1)^{i \alpha}S_{,j}Q_{\; \alpha,i}^{j} \delta \xi^{\alpha}.
\label{(110)}
\end{equation}

The action functionals (103), (104) may be expanded in functional Taylor
series about a background field. In generic notation one writes
\begin{equation}
S=S_{B}+(S_{,i})_{B}\phi^{i}+{1\over 2!}(S_{,ij})_{B}\phi^{j}\phi^{i}
+{1\over 3!}(S_{,ijk})_{B}\phi^{k}\phi^{j}\phi^{i}+...,
\label{(111)}
\end{equation}
$$
\phi^{i}=\varphi^{i}-\varphi_{B}^{\; i}.
$$
If the background fields satisfy the classical field equations then the 
second term on the right may be omitted.

The functional derivatives $(S_{,i_{1}...i_{N}})_{B}$ with 
$N \geq 3$ are known as (bare) {\bf vertex functions}. In the case of 
the Yang-Mills field the vertex functions with $N>4$ vanish and the
Taylor series terminates. In the case of the gravitational field the
series may or may not terminate depending on what choice is made for 
the basic field variables. By expressing inverse matrices in terms of
minors and determinants, and by examining the number of determinants 
needed to yield unit total weight for the integrand of (104), one easily
verifies that if the basic field variables are taken to be
${\cal G}^{\mu \nu} \equiv g^{r}g^{\mu \nu}$ and $r$ is chosen to be
$5/(4n+2)$ then the vertex functions with $N>2n+1$ vanish. 
Alternatively, if ${\cal G}_{\mu \nu}=g^{-r}g_{\mu \nu}$ are chosen
as the basic field variables, with $r=5/(6n-2)$, then the vertex
functions with $N > 3n-1$ vanish.\footnote{Both of these choices require
$n \not =2$. (See eqs. (40)).} There are three reasons, however, why 
neither of these choices is useful. First, the vertex functions of
gravity theory are exceedingly complicated, involving thousands of terms
already for $N=4$. Nobody is going to work out the vertex functions up to
maximum order even with the aid of a computer. Second, any imagined
advantage in these choices is lost as soon as one tries to introduce
dimensional regularization into the quantum theory. A specific choice
of field variables has then to be made, and it cannot vary continuously
with the dimension. Third, although a series that terminates has an
infinite radius of convergence, the range of the variables
$\phi^{\mu \nu} \equiv {\cal G}^{\mu \nu}-{\cal G}_{B}^{\; \mu \nu}$
or $\phi_{\mu \nu} \equiv {\cal G}_{\mu \nu}-{\cal G}_{B \mu \nu}$ is
in fact limited. These variables must avoid regions where the signature
of the metric tensor changes.

The third reason is the most important, at least in perturbation theory.
As is well known, the Feynman rules are obtained by inserting the
expansion (111) into the Feynman functional integral (see the next
section) and evaluating the integral as a sum (asymptotic series) of
Gaussian integrals, with the $\phi^{i}$ ranging from $-\infty$ to
$\infty$. Any constraint on the $\phi^{i}$ would make these integrals 
almost impossible to evaluate, and although one may for some purpose wish
to extend the Feynman integrand into nonphysical regions, one never does
this by naively removing constraints.

These remarks suggest, in fact, that none of the variables (40) is good
to use in perturbation theory. A better choice would be something like
\begin{equation}
\phi \equiv \Bigr[{\rm ln}(g \eta^{-1})\Bigr]\eta, \; 
g=e^{\phi \eta^{-1}} \eta, \;
\phi \equiv (\phi_{\mu \nu}), \; g \equiv (g_{\mu \nu}), \;
\eta \equiv (\eta_{\mu \nu}), \;
\eta^{-1} \equiv (\eta^{\mu \nu}),
\label{(112)}
\end{equation}
which maintains the signature of the spacetime metric. With these 
variables the series (111), of course, does not terminate, and one speaks
of gravity theory as being a {\bf non-polynomial Lagrangian theory}
(Isham, Strathdee and Salam (1971),(1972)). It will be noted that all
such ``safe'' variables inevitably transform {\bf nonlinearly} under
the diffeomorphism group.

Regardless of the choice of variables it is not difficult to draw
preliminary conclusions about the renormalizability or nonrenormalizability
(in perturbation theory) of a given quantum field theory. Although
momentum space is not, in an absolute sense, appropriate for use in
quantum gravity, conclusions about the high energy behaviour of amplitudes
in perturbation theory may be safely drawn with its aid. Consider a
Feynman graph with $L_{e}$ external lines, $L_{i}$ internal lines, and
$V_{N}$ $N$th-order vertices ($N \geq 3$). $L_{e},L_{i}$ and $V_{N}$
are related by the topological condition 
\begin{equation}
L_{e}+2L_{i}=\sum_{N}N V_{N}.
\label{(113)}
\end{equation}
The number of independent closed loops, or momentum integrations,
in the graph is given by
\begin{equation}
I=L_{i}-\sum_{N}V_{N}+1.
\label{(114)}
\end{equation}
In quantum gravity $2$ powers of momentum are associated with each
vertex, $-2$ powers with each internal line, and $n$ powers with each
momentum integration. The superficial degree of divergence of the graph
is therefore
\begin{equation}
D=-2L_{i}+2\sum_{N}V_{N}+nI=(n-2)I+2,
\label{(115)}
\end{equation}
which, for $n >2$, increases without limit as the number of independent
closed loops increases. This means that for $n>2$, there is an infinite
number of primitive divergences, and, if one attempts to compute order
by order, an infinite number of experimentally determined coupling 
constants is needed to determine the theory. These conclusions are
not altered if account is taken of the ``ghost'' contributions which,
as we shall see in the following sections, must be included. The theory
is said to be nonrenormalizable.

In Yang-Mills theory, in contrast, only one power of momentum is 
associated with each 3rd-order vertex, and the 4th-order vertices have
no momentum dependence at all. This leads to
\begin{equation}
D=-2L_{i}+\sum_{N}(4-N)V_{N}+nI=4+(n-4)I-L_{e}.
\label{(116)}
\end{equation}
For $n>4$ this theory too is nonrenormalizable, but for $n=4$ and an
arbitrary background field there are only four primitive divergences
(corresponding to $L=1,2,3,4$), and the theory is renormalizable. The
proof of renormalizability is not trivial and depends crucially on 
gauge invariance as well as some of the formal developments to be
discussed in the following sections. The primitive divergences turn out
to be related in virtue of gauge invariance.
 
The nonrenormalizability of standard quantum gravity has stimulated
investigations of alternative theories in which terms of the form
$g^{1/2}\Bigr(\alpha R^{2}+\beta R_{\mu \nu}R^{\mu \nu} \Bigr)$
are added to the integrand of expression (104). Such theories generally
suffer from ``physical ghosts'' with negative probabilities, but they do
improve the convergence situation. Each vertex now carries $4$ powers of
momentum and each internal line carries $-4$ powers. This leads to
\begin{equation}
D=-4 L_{i}+4 \sum_{N}V_{N}+nI=(n-4)I+4.
\label{(117)}
\end{equation}
When $n=4$ all diagrams have the same superficial degree of divergence, 
namely $4$. There is an infinity of primitive divergences, but they are
all related by gauge invariance, and only three experimental coupling
constants are required. The proof of renormalizability has been carried
out by Stelle (1977) using methods similar to those applied to
Yang-Mills theory.

It will be observed, in virtue of eq. (115), that the same methods 
should work in the case of standard quantum gravity when $n=2$,
although since the number of degrees of freedom in the field is then
{\bf negative} it is not clear what such a theory means. Weinberg
(1979) has studied the asymptotic stability of quantum gravity when
$n=2+\epsilon, \epsilon<<1$, and has given plausibility arguments
concerning its relevance for trying to make sense out of the theory 
when $n=4$.

In addition to its ultraviolet divergences quantum gravity also 
possesses infrared divergences. Gravitational field quanta 
-{\bf gravitons}- are massless. This fact in itself need not lead to
difficulties worse than those encountered in quantum electrodynamics 
where the divergences are completely understood and are removable by
standard methods. Gravitons, however, are coupled to other massless
quanta (photons, neutrinos, etc.) as well as to themselves. In
Yang-Mills theory as well as in massless electrodynamics such a
situation gives rise to infrared divergences of a new type that cannot
be removed by standard techniques or argued away on physical grounds.
In quantum gravity these new divergences are miraculously absent
(Weinberg (1965) and DeWitt (1967c)). It appears therefore that the
mysteries of Yang-Mills theory and gravity theory lie at opposite ends
of the momentum spectrum. There is an increasing body of evidence
that the Yang-Mills field solves its infrared dilemma by adopting a
nonstandard behaviour at long wavelengths, which is intimately related
to the phenomena of quark confinement and dynamical symmetry breaking.
These phenomena may also bear a technical relation to the failure of
gauge conditions to be globally valid in Yang-Mills theory (Gtibov (1977)).
No analogous phenomena are known to exist for gravity, at least when
spacetime is diffeomorphic to ${\bf R}^{n}$. The mysteries of gravitation
theory thus appear to lie solely at the high end of the momentum spectrum.

\section{The Feynman functional integral. Factoring out the gauge group.}

Consider a transition amplitude of the form $\langle {\rm out} | 
{\rm in} \rangle$ where the vectors $|{\rm in} \rangle$ and
$|{\rm out} \rangle$ refer to states in which the field is maximally
specified (in the quantum mechanical sense, e.g., in terms of complete
sets of commuting observables) in regions ``in'' and ``out'' 
respectively. These states need not be ``vacuum'' states and the regions
``in'' and ``out'' need not refer to the infinite past and future
respectively. If the background field (which enters naturally in most
calculations) has singularities in the past and/or future,
$| {\rm in} \rangle$ and $| {\rm out} \rangle$ may be defined not in 
terms of observables at all but by some analytic continuation procedure
(e.g., to the ``Euclidean sector'') that removes the singularities.
It will be assumed only that the ``in'' and ``out'' regions lie
respectively to the past and future of the region of dynamical interest.

There are many ways of showing that the amplitude 
$\langle {\rm out} | {\rm in} \rangle$ can be expressed as a formal
functional integral:
\begin{equation}
\langle {\rm out} | {\rm in} \rangle = N \int e^{iS[\varphi]}
\mu[\varphi] d\varphi, \; d\varphi \equiv \prod_{i}d\varphi^{i}.
\label{(118)}
\end{equation}
Here $N$ is a normalization constant, $S[\varphi]$ is the classical 
action functional, $\mu[\varphi]$ is chosen to make the 
``volume element'' $\mu d\varphi$ gauge invariant (see eq. (50)), and
the integration is to be extended over all fields $\varphi$ that satisfy
the boundary conditions appropriate to the given ``in'' and ``out'' 
states. We have remarked earlier (and will show later) that $\mu$ may be
set ``effectively'' equal to unity if the $\varphi^{i}$ are chosen to
transform linearly under the gauge group. We shall assume that such a
choice has been made and henceforth drop $\mu$ from the theory. (It can
always be restored if desired.)

Expression (118) was first derived by Feynman (1948) in ordinary quantum
mechanics, without gauge groups, and later (1950) applied by him to field
theory. The full extension to field theories with gauge groups is the
work of many people, and the student is referred to the literature for
details.\footnote{Useful modern references are Fadde'ev (1969), (1976) and
Abers and Lee (1973).}  When the Feynman integral is applied to the
gravitational field the only additional comment that needs to be made is
that the integration may have to embrace as many topologies as can be 
reached by analytic continuation from the given background topology.

If any of the fields $\varphi^{i}$ in (118) are fermionic the integration
with respect to them is to be carried out according to the formal rules
for integrating with respect to anticommuting variables that were first
introduced by Berezin (1966). These rules are analogous in many ways to the
well known rules for ordinary definite integrals from $-\infty$ to
$\infty$ with integrands that vanish asymptotically. For example, integrals
of total derivatives vanish, and the position of the zero point may be
shifted. On the other hand, with Berezin rules, transformations of 
variables and evaluation of Gaussian integrals lead to determinants 
precisely inverse to those of standard theory. When both bosonic and
fermionic fields are involved it is the {\bf super} determinant that
appears.
 
All physical amplitudes can be deduced from expression (118) by examining
how $\langle {\rm out} | {\rm in} \rangle$ changes under variations in
the action. Physical amplitudes can alternatively be obtained by judicious
use of
\begin{equation}
\langle {\rm out} | T(A[\varphi])| {\rm in} \rangle
=N \int A[\varphi] e^{i S[\varphi]} d\varphi,
\label{(119)}
\end{equation}
where $A[\phi]$ is any functional of the field {\bf operators} 
$\varphi^{i}$, and the $T$ symbol removes ambiguities about ordering 
the $\varphi^{i}$ by arranging them chronologically (with appropriate
$\pm$ signs thrown in if any of the $\varphi^{i}$ are fermionic).

When a gauge group is present the integration in (118) is redundant. 
Furthermore (119) is generally ambiguous, unless $A$ is gauge invariant,
in which case the integration in (119) too is redundant. This is
because, owing to the gauge invariance of the classical action, the
exponent in the integrands of (118) and (119) remains constant as $\varphi$
ranges over a group orbit in the configuration space $\Phi$. One can
remove this redundancy and/or ambiguity by adopting a gauge condition
like (94). The details of the procedure were first given by
Fadde'ev and Popov (1967).

Let $\xi$ be an element of the gauge group $G$, with coordinates 
$\xi^{\alpha}$, and let ${ }^{\xi}\varphi$ be the field to which
$\varphi$ is displaced under the action of $\xi$. Define
\begin{equation}
\bigtriangleup[\zeta,\varphi] \equiv \int_{G}
\delta[P[{ }^{\xi}\phi]-\zeta]{\rm det}Q^{-1}[\xi]d\xi, \;
d\xi \equiv \prod_{\alpha}d\xi^{\alpha},
\label{(120)}
\end{equation}
where $\delta[\;]$ is the delta {\bf functional}, $P^{\alpha}$ are the
functionals appearing in eq. (94), $Q^{-1}$ is the inverse of the matrix
formed out of the $Q_{\; \beta}^{\alpha}$ of eq. (53), and the integration
extends over the entire gauge group! We shall assume that the gauge
condition (94) is globally valid. The integrand in (120) then 
``switches on'' at only one point in $G$, namely that point for which
${ }^{\xi}\varphi$ is equal to the unique field $\varphi_{\zeta}$ lying
on the orbit containing $\varphi$ and picked out by the gauge condition:
\begin{equation}
P^{\alpha}[\varphi_{\zeta}]=\zeta^{\alpha}.
\label{(121)}
\end{equation}

By building infinitesimal parallelepipeds in the group manifold $G$ and
making use of eq. (53) one can verify that the combination
${\rm det}Q^{-1}[\xi]d\xi$ appearing in (120) is a right-invariant 
volume element, satisfying
\begin{equation}
{\rm det}Q^{-1}[\xi \xi']d(\xi \xi')={\rm det}Q^{-1}[\xi]d\xi \;
{\rm for} \; {\rm all} \; \xi' \; {\rm in} \; G.
\label{(122)}
\end{equation}
The presence of this volume element renders the functional 
gauge invariant:
\begin{equation}
\bigtriangleup[\zeta,{ }^{\xi'}\varphi]
=\bigtriangleup[\zeta,\varphi] \; {\rm for} \; {\rm all} \;
\xi' \; {\rm in} \; G.
\label{(123)}
\end{equation}
Several comments must be made about its use, however. In the case of the
diffeomorphism group, with the $Q$'s given by (54), it is easily checked
that
\begin{equation}
Q_{\; \; \; \; \; \nu'}^{-1 \mu}[\xi]
=\delta_{\; \nu}^{\mu} \delta(x,\xi(x'))
{\partial (\xi(x')) \over \partial (x')}.
\label{(124)}
\end{equation}
No one has ever discovered how to evaluate or give a meaning to the
determinant of this continuous matrix. The right-invariant volume element
of the diffeomorphism group, therefore, can only be defined (and used)
purely formally. The same is true for the invariance group of supergravity
theory. It should be noted that when the group is a supergauge group,
possessing anticommuting as well as commuting coordinates, 
det $Q^{-1}$ is a superdeterminant, and the integral (120) involves the
Berezin rules. (Remark: the delta functional in (120) presents no
difficulty. Delta functions of anticommuting variables turn out to be
easy to define. They can even be given Fourier representations.)

The gauge invariance of $\bigtriangleup$ makes it an easy functional to
evaluate. One has only to shift $\varphi$ to $\varphi_{\zeta}$ so that the
integrand in (120) switches on at the identity element $I$. All quantities
can then be expanded in power series in $\xi^{\alpha}-I^{\alpha}$. For
example, the argument of the delta functional becomes
\begin{eqnarray}
P^{\alpha}[{ }^{\xi}\varphi_{\zeta}]-\zeta^{\alpha}
&=& P^{\alpha}[\varphi_{\zeta}]-\zeta^{\alpha}
+P_{\; ,i}^{\alpha}[\varphi_{\zeta}]Q_{\; \beta}^{i}[\varphi_{\zeta}]
(\xi^{\beta}-I^{\beta})+... \nonumber \\
&=& F_{\; \beta}^{\alpha}[\varphi_{\zeta}](\xi^{\beta}+I^{\beta})+...
\label{(125)}
\end{eqnarray}
where 
\begin{equation}
F_{\; \beta}^{\alpha}[\varphi] \equiv P_{\; ,i}^{\alpha}[\varphi]
Q_{\; \beta}^{i}[\varphi].
\label{(126)}
\end{equation}
Similarly, making use of eq. (68), we find
\begin{equation}
Q_{\; \; \; \; \; \beta}^{-1 \alpha}[\xi]=\delta_{\; \beta}^{\alpha}
-Q_{\; \beta, \gamma}^{\alpha}[I](\xi^{\gamma}-I^{\gamma})+...
\label{(127)}
\end{equation}
and hence
\begin{eqnarray}
\bigtriangleup [\zeta,\varphi]&=& \int_{G}
\delta \Bigr[F[\varphi_{\zeta}](\xi-I)+... \Bigr]
\Bigr[1-(-1)^{\alpha}Q_{\; \alpha,\beta}^{\alpha}[I](\xi^{\beta}-I^{\beta})
+... \Bigr]d\xi \nonumber \\
&=& \Bigr({\rm det} F[\varphi_{\zeta}]\Bigr)^{-1},
\label{(128)}
\end{eqnarray}
$F$ being the matrix with elements $F_{\; \beta}^{\alpha}$. If any of the
group indices is fermionic the determinant is again a superdeterminant.
Note that if the $P$'s are constructed according to eq. (95),
$F[\varphi]$ is identical to the matrix $Q[P[\varphi]]$. Although this
construction will not be assumed in what follows, we shall, for simplicity
and convenience, assume that the gauge condition (94) is globally valid
for all $\zeta^{\alpha}$ lying in the ranges of the functionals 
$P^{\alpha}[\varphi]$.

The next step is to insert unity into the integrand of (118), in the
guise of 
$$
(\bigtriangleup[\zeta,\varphi])^{-1} \int_{G}
\delta \Bigr[P[{ }^{\xi}\varphi]-\zeta \Bigr]{\rm det}Q^{-1}[\xi]d\xi,
$$
and interchange the order of integrations, obtaining
\begin{equation}
\langle {\rm out} | {\rm in} \rangle = N \int_{G} {\rm det} Q^{-1}[\xi]
d\xi \int d\varphi e^{iS [\varphi]}
(\bigtriangleup[\zeta,\varphi])^{-1}
\delta \Bigr[P[{ }^{\xi}\varphi]-\zeta \Bigr].
\label{(129)}
\end{equation}
We have assumed a choice of variables for which the volume element
$d\varphi$ is gauge invariant ($\mu=1$). $S[\varphi]$ and
$\bigtriangleup[\zeta,\varphi]$ are also gauge invariant. Therefore a
superscript $\xi$ may be affixed to every $\varphi$ in the integrand
of (129) that does not already bear one. But every 
${ }^{\xi}\varphi$ is then a dummy, and hence all the $\xi$'s may be
removed. Making use of (128) one immediately obtains
\begin{equation}
\langle {\rm out} | {\rm in} \rangle =N' \int e^{i S[\varphi]}
{\rm det} F[\varphi] \delta \Bigr[P[\varphi]-\zeta \Bigr],
\label{(130)}
\end{equation}
where $F[\varphi_{\zeta}]$ has been replaced by $F[\varphi]$ in the
integrand because of the presence of the delta functional, and where
\begin{equation}
N' \equiv N \int_{G} {\rm det}Q^{-1}[\xi]d\xi.
\label{(131)}
\end{equation}

The gauge group has now been factored out, and its ``volume'' has been
absorbed into the new normalization constant $N'$. The integration in
(130) is restricted to the subspace $P^{\alpha}[\varphi]=\zeta^{\alpha}$.

The technique of confining the fields $\varphi^{i}$ to a
particular subspace can also be used to remove the ambiguity from the
integral (119) when $A[\varphi]$ is not gauge invariant. Strictly speaking,
matrix elements are definable only for gauge invariant operators. 
However, given a non-gauge-invariant operator 
$A[{\underline \varphi}]$, one can construct
a gauge invariant operator out of it by the following definition:
\begin{equation}
T(A[{\underline \varphi}_{\zeta}]) \equiv 
T \left( (\bigtriangleup[\zeta,{\underline \varphi}])^{-1}
\int_{G} A[{ }^{\xi}{\underline \varphi}] 
\delta \Bigr[P[{ }^{\xi}{\underline \varphi}]-\zeta \Bigr]
{\rm det}Q^{-1}[\xi] d\xi. \right)
\label{(132)}
\end{equation}
The chronological ordering symbol is used here so that the non-commutativity
(or anti-commutativity) of $A[{ }^{\xi}{\underline \varphi}]$ with both 
$(\bigtriangleup[\zeta,{\underline \varphi}])^{-1}$ 
and the delta functional can be
effectively ignored. Note that because the gauge group acts linearly on
the ${\underline \varphi}$'s there is no ambiguity about the symbol
${ }^{\xi}{\underline \varphi}$. Note, however, 
that diffeomorphisms in gravity theory
can drag the field in very complicated ways. The chronological operation,
which orders field operators solely by the value of the coordinate 
$x^{0}$, rearranges the ``physical'' fields in correspondingly complicated
ways as the variable $\xi$ in the integral (132) ranges over the group.

Applying eq. (119) to the operator 
$T(A[{\underline \varphi}_{\zeta}])$ and following
the same reasoning as was used in passing from eq. (129) to eq. (130),
one finds
\begin{equation}
\langle {\rm out} | T(A[{\underline \varphi}_{\zeta}])| 
{\rm in} \rangle
=N' \int A[\varphi] e^{iS[\varphi]} {\rm det}F[\varphi]
\delta[P[\varphi]-\zeta]d\varphi,
\label{(133)}
\end{equation}
valid for any functional $A[\varphi]$. 

\section{Averaging over gauges}

It is possible to develop a perturbation theory based on eqs. (130)
and (133), but it is usually more convenient to work with a formalism
from which the delta functionals have been eliminated. Note that although
the parameters $\zeta^{\alpha}$ appear on the right side of (130), the
amplitude $\langle {\rm out} | {\rm in} \rangle$ is actually independent
of them. Therefore nothing changes if we integrate over these parameters,
with a weight factor.

In practically all studies of non-Abelian gauge theories to date, 
Gaussian weight factors of the form 
$$
{\rm exp} \left({1\over 2}i \zeta^{\alpha} \; { }_{\alpha}M_{\beta}
\; \zeta^{\beta}\right),
$$ 
where $M$ is a nonsingular constant matrix having the symmetry 
$$
{ }_{\alpha}M_{\beta}=(-1)^{\alpha+\beta+\alpha \beta}
{ }_{\beta}M_{\alpha},
$$
have been used. From a fundamental standpoint a Gaussian weight factor
can be used only if the bosonic $\zeta$'s can range from $-\infty$
to $\infty$ without the gauge condition (94) becoming globally invalid -
for example, if the $P$'s satisfy eq. (95), with $Q_{\; \beta}^{\alpha}$'s
based on canonical group coordinates (eq. (69)). This condition, however,
is almost universally violated. Indeed, the Gaussian weight function is
most frequently employed in combination with linear gauge conditions like
(98), where it almost certainly introduces errors globally (i.e., in
non-perturbative analyses).

In the case of quantum gravity, where the gauge group has no canonical
coordinates, it seems particularly inappropriate to confine our attention
to Gaussian weight factors. We shall therefore introduce a more general
weight factor, of the form ${\rm exp}(iU[\zeta])$, where we specify nothing
about the functional $U[\zeta]$ except the following three conditions:
\vskip 0.3cm
\noindent
(1) $U$ becomes infinite on the boundary of the allowable domain of the
$\zeta$'s (which domain we are assuming coincides with the range of the
$P[\varphi]$'s). 
\vskip 0.3cm
\noindent
(2) $U$ and all its first functional derivatives $U_{,\alpha}$ vanish
at some chosen point (e.g., at $\zeta^{\alpha}=I^{\alpha}$ when the
$P$'s are group coordinates); its second functional derivatives 
$U_{,\alpha \beta}$, however, form a nonsingular continuous matrix at
that point. 
\vskip 0.3cm
\noindent
(3) $U$ vanishes nowhere else, and its first derivatives all vanish
simultaneously nowhere else. 
\vskip 0.3cm
The third condition is imposed mainly for convenience. Note that all
three are satisfied by the Gaussian exponent 
${1\over 2}\zeta^{\alpha} \; { }_{\alpha}M_{\beta} \; \zeta^{\beta}$ 
whenever it can be legitimately used. 

Inserting ${\rm exp}(iU[\varphi])$ into the integrand of eq. (130)
and integrating over the $\zeta$'s, one obtains
\begin{equation}
\langle {\rm out} | {\rm in} \rangle = N''[U]
\int e^{i(S[\varphi]+U[P[\varphi]])}
{\rm det} F[\varphi] d\varphi,
\label{(134)}
\end{equation}
\begin{equation}
N''[U] \equiv {N' \over \int e^{iU[\zeta]} d\zeta},
\label{(135)}
\end{equation}
the integration domain in (135) being understood to be the allowable
domain of the $\zeta$'s. Equation (133) too may be replaced by a weighted
average. Defining
\begin{equation}
T(A[\varphi]) \equiv \Bigr(\int e^{iU[\zeta]}d\zeta \Bigr)^{-1}
\int T(A[\varphi_{\zeta}])e^{iU[\zeta]}d\zeta,
\label{(136)}
\end{equation}
one may write
\begin{equation}
\langle {\rm out} | T(A[\varphi])|{\rm in} \rangle = N''[U]
\int A[\varphi] e^{i(S[\varphi]+U[P[\varphi]])}
{\rm det}F[\varphi] d\varphi.
\label{(137)}
\end{equation}
Equation (137), and generalizations of it, will be used frequently
in the following sections. Definitions (132) and (136) reveal precisely 
what kind of averaged quantum operator is associated with each classical
functional $A[\varphi]$ in this formalism. Note that if $f[\zeta]$ is
any functional of the $\zeta$'s we have
\begin{equation}
T(f[P[\varphi_{\zeta}]])=f[\zeta]
\label{(138)}
\end{equation}
and
\begin{equation}
\langle {\rm out} | T(f[P[\varphi]])| {\rm in} \rangle 
= \langle f \rangle \langle {\rm out} | {\rm in} \rangle 
\label{(139)}
\end{equation}
where
\begin{equation}
\langle f \rangle \equiv {\int f[\zeta]e^{iU[\zeta]}d\zeta \over
\int e^{iU[\zeta]} d\zeta}.
\label{(140)}
\end{equation}

Having so freely manipulated formal expressions we should now check that
no inconsistencies have crept into our results, by verifying directly
that the right side of (134), for example, is truly independent of the
choices we have made for the functionals $P^{\alpha}[\varphi]$ and
$U[\zeta]$. Obviously the right side {\bf will} be affected if we naively
switch to $P$'s for which the gauge condition (94) is no longer globally
valid for all $\zeta$'s in the range of the $P$'s. Therefore we must
assume that the changes $\delta P^{\alpha}$ (which, without loss of 
generality, may be taken infinitesimal) maintain global validity.

We also confine our attention to changes $\delta U$ that leave the
location of the zero of $U$, as well as the three conditions that
we imposed upon $U$, intact. It is not difficult to see that $\delta U$
may then always be expressed in the form
\begin{equation}
\delta U[\zeta]=U_{,\alpha}[\zeta] \delta V^{\alpha}[\zeta],
\label{(141)}
\end{equation}
where the $\delta V^{\alpha}$ vanish at the zero of $U$. Note that
under this change we have
\begin{equation}
\delta N''[U]=-iN''[U]{\int e^{iU[\zeta]}U_{,\alpha}[\zeta]
\delta V^{\alpha}[\zeta]d\zeta \over 
\int e^{iU[\zeta]} d\zeta }
=N''[U]\langle (-1)^{\alpha}\delta V_{\; ,\alpha}^{\alpha} \rangle,
\label{(142)}
\end{equation}
the final form being obtained by an integration by parts in which the
boundary of the integration domain contributes nothing because 
${\rm exp}(iU[\zeta])$ oscillates infinitely rapidly there.

Making use of eqs. (126), (139), (141) and (142) we now have
\begin{eqnarray}
\delta \langle {\rm out} | {\rm in} \rangle &=& N''[U]
\int e^{i(S[\varphi]+U[P[\varphi]])} \Bigr \{ (-1)^{\alpha}
\delta V_{\; ,\alpha}^{\alpha} [P[\varphi]] \nonumber \\
&+& i U_{,\alpha}[P[\varphi]]\Bigr(\delta V^{\alpha}[P[\varphi]]
+\delta P^{\alpha}[\varphi]\Bigr) \nonumber \\
&+& (-1)^{\alpha}F_{\; \; \; \; \; \; \beta}^{-1 \alpha}[\varphi]
\delta P_{\; ,i}^{\beta}[\varphi] Q_{\; \alpha}^{i}[\varphi]
\Bigr \} {\rm det}F[\varphi]d\varphi,
\label{(143)}
\end{eqnarray}
where the inverse $F^{-1}$, if it is a Green's function (as it often
will be), must satisfy the boundary conditions appropriate to the
``in'' and ``out'' states. The integral (143) does not obviously vanish.
The way to show that it is nevertheless zero is as follows. Replace
each $\varphi^{i}$ in the integral (134) by ${\overline \varphi}^{i}$,
where
\begin{equation}
{\overline \varphi}^{i}=\varphi^{i}+Q_{\; \alpha}^{i}[\varphi]
\delta \xi^{\alpha}[\varphi],
\label{(144)}
\end{equation}
\begin{equation}
\delta \xi^{\alpha}[\varphi]=F_{\; \; \; \; \; \; \beta}^{-1 \alpha}[\varphi]
\Bigr(\delta V^{\beta}[P[\varphi]]+\delta P^{\beta}[\varphi]\Bigr).
\label{(145)}
\end{equation}
Since the $\varphi$'s are just dummies this replacement has no effect. 
However, it is not difficult to show that the net apparent change in the
integral is given precisely by (143) {\bf provided} one is entitled to make
the identifications
\begin{equation}
(-1)^{i(\alpha+1)} Q_{\; \alpha,i}^{i}=0, \;
(-1)^{\beta(\alpha+1)}C_{\; \alpha \beta}^{\beta}=0.
\label{(146)}
\end{equation}
We shall comment on these equations presently.

It is easy to see that the second term inside the curly brackets in 
(143) comes from the change that the replacement 
$\varphi \rightarrow {\overline \varphi}$ induces in the exponent of
(134). That the first and third terms come from the change in the 
product ${\rm det}F[\varphi]d\varphi$ may be shown as follows. 
First compute
$$
{\overline \phi}_{\; ,j}^{i}=\delta_{\; j}^{i}
+(-1)^{j \alpha} Q_{\; \alpha,j}^{i}[\varphi]
\delta \xi^{\alpha}[\varphi] 
$$
$$
-Q_{\; \alpha}^{i}[\varphi]F_{\; \; \; \; \; \; \beta}^{-1 \alpha}[\varphi]
\Bigr((-1)^{jk}P_{\; ,kj}^{\beta}[\varphi]Q_{\; \gamma}^{k}[\varphi]
+(-1)^{j \gamma}P_{\; ,k}^{\beta}[\varphi]
Q_{\; \gamma ,j}^{k}[\varphi]\Bigr) \delta \xi^{\gamma}[\varphi]
$$
$$
+Q_{\; \alpha}^{i}[\varphi]F_{\; \; \; \; \; \; \beta}^{-1 \alpha}[\varphi]
\Bigr(\delta V_{\; ,\gamma}^{\beta}[P[\varphi]]
P_{\; ,j}^{\gamma}[\varphi]+\delta P_{\; ,j}^{\beta}[\varphi]\Bigr),
$$
which, after rearrangement of some factors and use of eq. (126),
yields the (super) Jacobian
\begin{eqnarray}
{\rm det} \Bigr({\overline \phi}_{\; ,j}^{i}\Bigr)
&=& 1+(-1)^{i(\alpha+1)}Q_{\; \alpha,i}^{i}[\varphi]
\delta \xi^{\alpha}[\varphi] \nonumber \\
&-& F_{\; \; \; \; \; \; \beta}^{-1 \alpha}[\varphi]\biggr((-1)^{\alpha (j+1)}
P_{\; ,ji}^{\beta}[\varphi]Q_{\; \alpha}^{i}[\varphi]
Q_{\; \gamma}^{j}[\varphi] \nonumber \\
&+& (-1)^{\alpha(\gamma+1)}P_{\; ,j}^{\beta}[\varphi]
Q_{\; \gamma,i}^{j}[\varphi] Q_{\; \alpha}^{i}[\varphi]\biggr)
\delta \xi^{\gamma}(\varphi) \nonumber \\
&+& (-1)^{\alpha}\delta V_{\; ,\alpha}^{\alpha}[P[\varphi]]
+(-1)^{\alpha}F_{\; \; \; \; \; \; \beta}^{-1 \alpha}[\varphi]
\delta P_{\; ,i}^{\beta}[\varphi]Q_{\; \alpha}^{i}[\varphi].
\label{(147)}
\end{eqnarray}
Combining $\delta d\varphi \equiv d{\overline \varphi}-d\varphi
=\Bigr[{\rm det}\Bigr({\overline \varphi}_{\; ,j}^{i}\Bigr)-1 \Bigr]$ 
with
$$
\delta {\rm det}F[\varphi \equiv {\rm det}F[{\overline \varphi}]
-{\rm det}F[\varphi]
$$
$$
=(-1)^{\alpha}{\rm det}F[\varphi]
F_{\; \; \; \; \; \; \beta}^{-1 \alpha}[\varphi]
F_{\; \alpha,i}^{\beta}[\varphi] Q_{\; \gamma}^{i}[\varphi]
\delta \xi^{\gamma}[\varphi]
$$
$$
=(-1)^{\alpha}{\rm det}F[\varphi]
F_{\; \; \; \; \; \; \beta}^{-1 \alpha}[\varphi]
\biggr((-1)^{\alpha i}P_{\; ,ij}^{\beta}[\varphi]
Q_{\; \alpha}^{j}[\varphi] Q_{\; \gamma}^{i}[\varphi]
$$
$$
+P_{\; ,j}^{\beta}[\varphi]Q_{\; \alpha,i}^{j}[\varphi]
Q_{\; \gamma}^{i}[\varphi]\biggr)\delta \xi^{\gamma}[\varphi],
$$
and making use of eq. (29), one finds for the change in
${\rm det}F[\varphi]d\varphi$ under the replacement
$\varphi \rightarrow {\overline \varphi}$,
\begin{eqnarray}
\; & \; & \delta \Bigr({\rm det}F[\varphi]d\varphi \Bigr) \nonumber \\
&=& \biggr \{ \Bigr[(-1)^{i(\alpha+1)}Q_{\; \alpha ,i}^{i}
-(-1)^{\beta (\alpha+1)}C_{\; \alpha \beta}^{\beta}\Bigr]
\delta \xi^{\alpha}[\varphi] \nonumber \\
&+& (-1)^{\alpha}\delta V_{\; ,\alpha}^{\alpha}[P[\varphi]]
+(-1)^{\alpha}F_{\; \; \; \; \; \; \beta}^{-1 \alpha}[\varphi]
\delta P_{\; ,i}^{\beta}[\varphi]Q_{\; \alpha}^{i}[\varphi]
\biggr \} {\rm det}F[\varphi]d\varphi.
\label{(148)}
\end{eqnarray}
If eqs. (146) are assumed to hold one is left with precisely the first
and third terms inside the curly brackets in (143).

Equations (146) were not needed in the derivation of eq. (134). Why are
they needed now? When the gauge group has no anticommuting coordinates
the answer is that eqs. (146) are forced on us by the procedure of
factoring out the gauge group. Our interchanging the orders of integration
in arriving at eq. (129), and our use of eq. (131), amount to adopting
the rule that the gauge group is to be treated formally as if it were
{\bf compact}. For consistency the associated Lie algebra must likewise 
be treated as compact. The generators of real representations of compact
Lie algebras all have vanishing trace. Hence eqs. (146). (Remember, we
are assuming that the $\varphi$'s transform linearly under the
gauge group.)

In Yang-Mills theories eqs. (146) hold automatically because the generating
group is always compact. In gravity theory the situation is more subtle.
Both $Q_{\; \alpha,i}^{i}$ and $C_{\; \alpha \beta}^{\beta}$, if one
tries to compute them from eqs. (28) and (35), are meaningless expressions
involving derivatives of delta functions with coincident arguments. 
However, both are metric-independent covariant vector densities of unit
weight. Any sensible regularization scheme {\bf must} assign them the
value zero, for otherwise spacetime would be endowed with a preferred 
direction even before a metric is imposed on it.

If $G$ is a {\bf super}gauge group, with anticommuting coordinates, the
formal compactness argument fails. But the notion of simplicity, or
semisimplicity, survives. The invariance groups of all known supergauge
theories are semisimple, and the generators of real representations of 
such groups satisfy the supertrace laws (146). The semisimplicity
argument can also be invoked in the case of the local frame group, which
enters when the gravitational field is expressed in terms of local frame
components rather than directly in terms of the metric tensor (e.g., when
spinor fields are present).

If we choose $\varphi$'s that do not transform linearly under the group
then the functional $\mu[\varphi]$ of eq. (118) has to be reintroduced
into the theory. It is easy to verify that consistency of the above
formalism is maintained under these circumstances provided the first of
eqs. (146) is replaced by eq. (50), which is just the condition that the
product $\mu[\varphi]d\varphi$ be gauge invariant. Equation (50) is, of
course, consistent with the first of eqs. (146) when $\mu=1$.

At this point the student may object that, in the case of quantum gravity
at least, there appears to be an inconsistency in what we have done. 
Consider the sets of variables defined by eqs. (40). {\bf All} of these
sets transform linearly under the diffeomorphism group. However, the
Jacobian that arises in transforming from one set to another is not 
generally constant. How can one maintain $\mu$ = constant for all sets?
The answer is that one {\bf must}. If the Jacobian is replaced by the
exponential of its logarithm, it contributes a formally divergent term of
the form ${\rm const}. \times \delta(0) \int {\rm ln}g d^{n}x$ to the
exponents in the Feynman functional integrals. All terms of this kind
must be suppressed by any viable regularization scheme. By this criterion
the dimensional regularization method, for example, is a viable scheme.

\section{Ghosts. The BRS transformation. The generating functional}

The perturbation rules to which eqs. (134) and (137) lead may be 
summarized as follows. The exponent in the integrands is, as usual,
expanded about a stationary background $\varphi_{B}$, and the integrals
are evaluated as series of Gaussian integrals. If ${\rm exp}(iU)$ is a
Gaussian weight factor and the $P^{\alpha}$ are chosen (unwisely) to
have the linear form (98), then the vertex functions are just the
functional derivatives $S_{,i_{1}...i_{N}}, \; N \geq 3$. Otherwise the
vertex functions include contributions from $U[P[\varphi]]$. In addition
to the usual graphs that one can draw there is an infinite set of new
graphs arising from the factor ${\rm det}F[\varphi]$, involving a new set
of ``formal particles'' called {\bf ghosts}. The inverse matrix 
$F_{\; \; \; \; \; \; \beta}^{-1 \alpha}[\varphi]$ 
is the bare ghost propagator 
in an arbitrary field $\varphi$, i.e., with an arbitrary number of
$\varphi$-lines attached. The ghost propagators always enter in closed
loops, never as external lines.

The conditions previously imposed on the functional $U$ ensure that the
$\varphi$-propagator exists. The presence of $U[P[\varphi]]$ in the
exponents of expressions (134) and (137) breaks the gauge symmetry and
eliminates the redundancy that exists in the integration (118). An
important symmetry nevertheless survives. It is most easily revealed by
introducing two new fields, $\chi_{\alpha}$ and $\psi^{\alpha}$, that
have the unusual property of being fermionic when the index $\alpha$
is bosonic and vice versa. Use of these fields together with the Berezin
integration rules allows one to express ${\rm det}F[\varphi]$ in the form
\begin{equation}
\int e^{i \chi_{\alpha} F_{\; \beta}^{\alpha}[\varphi] \psi^{\beta}}
d\chi \; d\psi=C \; {\rm det}F[\varphi],
\label{(149)}
\end{equation}
where $C$ is a (divergent) constant, and hence
\begin{equation}
\langle {\rm out} | {\rm in} \rangle ={\overline N}[U]
\int e^{i(S[\varphi]+U[P[\varphi]]+\chi F[\varphi] \psi)}
d\varphi \; d\chi \; d\psi
\label{(150)}
\end{equation}
\begin{equation}
{\overline N}[U] \equiv N''[U]/C.
\label{(151)}
\end{equation}
Equation (150) shows that the fields $\chi_{\alpha},\psi^{\alpha}$ are
associated with the ghost particles, which are now placed on a common
footing with the $\varphi$-particles.

It was discovered by Becchi, Rouet and Stora (BRS) (1975) that both the
exponent and the volume element $d\varphi \; d\chi \; d\psi$ in (150) are
invariant under a set of transformations whose infinitesimal forms are
given by
\begin{equation}
d\varphi^{i}=Q_{\; \alpha}^{i}[\varphi]\psi^{\alpha} \; \delta \lambda, \;
\delta \chi_{\alpha}=\delta \lambda \; U_{,\alpha}[P[\varphi]], \;
\delta \psi^{\alpha}=-{1\over 2}C_{\; \beta \gamma}^{\alpha}
\psi^{\gamma} \; \delta \lambda \; \psi^{\beta},
\label{(152)}
\end{equation}
where $\delta \lambda$ is an arbitrary infinitesimal anticommuting 
constant. Using the special (anti)commutativity properties of the
$\chi$'s and $\psi$'s, together with the identity (29) and the definition 
(31), one readily verifies the invariance of the exponent. By computing
the super-Jacobian of the BRS transformation one finds that the volume
element $d\varphi \; d\chi \; d\psi$ is likewise invariant, provided
eqs. (146) are assumed to hold. It is also straightforward to verify
that, if confined to the $\varphi$'s and $\psi$'s, the BRS transformations
constitute an Abelian group. Inclusion of the $\chi$'s destroys the
group property unless $F_{\; \beta}^{\alpha}[\varphi]\psi^{\beta}=0$.
Note that the BRS transformations do {\bf not} constitute a local gauge
group. The $\delta \lambda$'s are constants; they are not functions over
spacetime. Thus the integral (150) contains no redundancy.

The BRS transformations play an important role in simplifying the
derivation of the ``Ward-Takahashi identity'' satisfied by the
so-called generating functional. What follows is a partial account, adapted
to the case in which the $P$'s are nonlinear and ${\rm exp}(iU)$ is
non-Gaussian, of the theory of the generating functional given by
B.W. Lee (1976) and originally due to Zinn-Justin.

One begins by replacing the exponent in eq. (150) by
$$
{\widetilde S}[\varphi,\chi,\psi,K,L,M]+J_{i}\varphi^{i}
+{\overline J}^{\alpha}\chi_{\alpha}+{\widehat J}_{\alpha}\psi^{\alpha},
$$
where
\begin{eqnarray}
{\widetilde S}[\varphi,\chi,\psi,K,L,M] & \equiv & S[\varphi]
+U[P[\varphi]]+\chi_{\alpha} F_{\; \beta}^{\alpha}[\varphi]\psi^{\beta}
\nonumber \\
&+& \left \{ K_{i}+M(U[P[\varphi]])_{,i} \right \}
Q_{\; \alpha}^{i}[\varphi]\psi^{\alpha}
-{1\over 2}(-1)^{\beta}L_{\alpha}C_{\; \; \beta \gamma}^{\alpha}
\psi^{\gamma}\psi^{\beta}
\label{(153)}
\end{eqnarray}
and by generalizing eq. (137) to
\begin{equation}
\langle {\rm out} | T(A[\varphi,\chi,\psi])| {\rm in} \rangle \equiv
{\overline N}[U] \int A[\varphi,\chi,\psi]
e^{i({\widetilde S}+J \varphi +{\overline J}\chi
+{\widehat J}\psi)}d\varphi \; d\chi \; d\psi.
\label{(154)}
\end{equation}
$J_{i},{\overline J}^{\alpha},{\widehat J}_{\alpha},K_{i},L_{\alpha}$
and $M$ are {\bf external sources}\footnote{$M$ is a constant. The others
depend, through their indices, on position in spacetime.}, and the
``matrix element'' (154) is a functional of them. $J_{i}$ and $L_{\alpha}$
are bosonic when their indices are bosonic and fermionic when their 
indices are fermionic. With ${\overline J}^{\alpha}, {\widehat J}_{\alpha}$
and $K_{i}$ the association is just the opposite. $M$ is fermionic.

If the functional $A$ in (154) is replaced by unity one gets a
generalization of the ``in-out'' amplitude:
\begin{eqnarray}
\; & \; & e^{iW[J,{\overline J},{\widehat J},K,L,M]} \equiv 
\langle {\rm out} | {\rm in} \rangle \nonumber \\
& \equiv & {\overline N}[U]\int e^{i({\widetilde S}+J \varphi
+{\overline J}\chi +{\widehat J}\psi)}
d\varphi \; d\chi \; d\psi.
\label{(155)}
\end{eqnarray}
This generalized amplitude is called the {\bf generating functional},
because if it is expanded in a power series in the sources 
$J_{i},{\overline J}^{\alpha}$ and ${\widehat J}_{\alpha}$ the coefficients
are the matrix elements of chronological products of field operators.
The coefficient of zero order reduces to the original amplitude (150) when
$K_{i},L_{\alpha}$ and $M$ vanish.

The functional ${\widetilde S}$ may be viewed as a generalized action 
functional. With the aid of eqs. (29) and (36) one may readily show that
it is BRS invariant. Suppose the variables $\varphi,\chi,\psi$ in the
integrand of (155), as well as in the volume element, are subjected to a
BRS transformation. Since these variables are dummies the integral remains
unaffected. Explicitly, however, the terms in 
$J,{\overline J},{\widehat J}$ change. Therefore
\begin{eqnarray}
\; & \; & 0=i {\overline N}[U] \int \left \{ J_{i} Q_{\; \alpha}^{i}[\varphi]
\psi^{\alpha}+(-1)^{\alpha}{\overline J}^{\alpha}U_{,\alpha}[P[\varphi]]
+{1\over 2}(-1)^{\beta}{\widehat J}_{\alpha}C_{\; \beta \gamma}^{\alpha}
\psi^{\gamma}\psi^{\beta} \right \} \nonumber \\
& \times & e^{i({\widetilde S}+J \varphi +{\overline J}\chi
+{\widehat J}\psi)}d\varphi \; d\chi \; d\psi.
\label{(156)}
\end{eqnarray}
This result can be expressed in an alternative form through use of  
\begin{eqnarray}
0 &=& \int {\delta \over \delta \chi_{\alpha}}\left \{ f[\varphi]
e^{i({\widetilde S}+J \varphi +{\overline J}\chi+{\widehat J}\psi)} \right \}
d\varphi \; d\chi \; d\psi \nonumber \\
&=& i \int \left \{ F_{\; \beta}^{\alpha}[\varphi]\psi^{\beta}
-(-1)^{\alpha}{\overline J}^{\alpha}\right \} f[\varphi]
e^{i({\widetilde S}+J \varphi +{\overline J}\chi
+{\widehat J}\psi)} d\varphi \; d\chi \; d\psi,
\label{(157)}
\end{eqnarray}
where $f$ is any functional of the $\varphi$'s. One obtains
\begin{eqnarray}
0&=& i {\overline N}[U] \int \left \{ J_{i} 
Q_{\; \alpha}^{i}[\varphi] \psi^{\alpha}
+U_{,\alpha}[P[\varphi]]F_{\; \beta}^{\alpha}[\varphi]\psi^{\beta}
+{1\over 2}(-1)^{\beta}{\widehat J}_{\alpha} C_{\; \beta \gamma}^{\alpha}
\psi^{\gamma} \psi^{\beta} \right \} \nonumber \\
& \times & e^{i({\widetilde S}+J \varphi +{\overline J}\chi 
+{\widehat J}\psi)} d\varphi \; d\chi \; d\psi \nonumber \\
&=& {\overline N}[U] \int \left(J_{i}{\delta \over \delta K_{i}}
+{\partial \over \partial M}-{\widehat J}_{\alpha}
{\delta \over \delta L_{\alpha}}\right)
e^{i({\widetilde S}+J \varphi +{\overline J}\chi
+{\widehat J}\psi)} d\varphi \; d\chi \; d\psi,
\label{(158)}
\end{eqnarray}
in which use is made of the fact that the term containing $M$ in 
$\widetilde S$ may be written in the form
$$
M U_{,\alpha}[P[\varphi]]F_{\; \beta}^{\alpha}[\varphi]\psi^{\beta}.
$$
Multiplying eq. (158) by $-ie^{-iW}$, we finally get
\begin{eqnarray}
0&=& -ie^{-iW} \left(J_{i}{\delta \over \delta K_{i}}
+{\partial \over \partial M}-{\widehat J}_{\alpha}
{\delta \over \delta L_{\alpha}}\right)e^{iW} \nonumber \\
&=& J_{i}{\delta W \over \delta K_{i}}+{\partial W \over \partial M}
-{\widehat J}_{\alpha}{\delta W \over \delta L_{\alpha}}.
\label{(159)}
\end{eqnarray}
This relation expresses an important symmetry property of the generating
functional, which leads directly to the Ward-Takahashi identities to be
discussed presently. First we must review some standard material on the
so-called effective action.

\section{Many-particle Green's functions. The effective action}

In this section important use will be made of the {\bf Schwinger average}:
\begin{equation}
\langle {\underline A} \rangle \equiv {\langle {\rm out} 
| T({\underline A}) | {\rm in} \rangle
\over \langle {\rm out} | {\rm in} \rangle }.
\label{(160)}
\end{equation}
Here ${\underline A}$ is an arbitrary functional of the operators
${\underline \varphi}^{i},{\underline \chi}_{\alpha},
{\underline \psi}^{\alpha}$, and the numerator and
denominator on the right are defined by eqs. (154) and (155) respectively.
It will be convenient to define
\begin{equation}
\varphi^{i} \equiv \langle {\underline \varphi}^{i} \rangle, \;
\chi_{\alpha} \equiv \langle {\underline \chi}_{\alpha} \rangle , \;
\psi^{\alpha} \equiv \langle {\underline \psi}^{\alpha} \rangle.
\label{(161)}
\end{equation}
When the sources ${\overline J}^{\alpha},{\widehat J}_{\alpha},K_{i},
L_{\alpha},M$ vanish, the averages $\chi_{\alpha}$ and $\psi^{\alpha}$ vanish. 
Note that although the symbols $\varphi^{i},\chi_{\alpha},\psi^{\alpha}$
have previously been used for integration variables, no confusion about
their meaning will arise in practice. 

It will also be convenient to denote the operators 
${\underline \varphi}^{i},{\underline \chi}_{\alpha}$, and
${\underline \psi}^{\alpha}$ collectively by ${\underline \varphi}^{A}$,
their averages by $\varphi^{A}$, and the sources 
$J_{i},{\overline J}^{\alpha}$ and ${\widehat J}_{\alpha}$ collectively
by $J_{A}$. Let $\bigtriangleup J_{A}$ be arbitrary finite increments in 
the sources. Then we may write
\begin{eqnarray}
\; & \; & \sum_{n=0}^{\infty}{i^{n}\over n!}
\bigtriangleup J_{A_{n}}... \bigtriangleup J_{A_{1}}
\langle {\rm out} | T({\underline \phi}^{A_{1}}...
{\underline \phi}^{A_{n}})| {\rm in} \rangle \nonumber \\
&=& {\rm exp} \left(\bigtriangleup J_{A} {\delta \over \delta J_{A}}
\right) \langle {\rm out} | {\rm in} \rangle 
=\Bigr(e^{iW}\Bigr)_{J \rightarrow J + \bigtriangleup J}
\nonumber \\
&=& {\rm exp} \left(iW +i \bigtriangleup J_{A}\varphi^{A}
+i \sum_{n=2}^{\infty}{1\over n!} \bigtriangleup J_{A_{n}}
... \bigtriangleup J_{A_{1}}G^{A_{1}...A_{n}}\right),
\label{(162)}
\end{eqnarray}
where 
\begin{equation}
\varphi^{A}=\langle {\underline \varphi}^{A} \rangle
=e^{-iW} {\delta \over i \delta J_{A}}e^{iW}
={\delta W \over \delta J_{A}},
\label{(163)}
\end{equation}
\begin{equation}
G^{A_{1}...A_{n}} \equiv {\delta \over \delta J_{A_{1}}} ...
{\delta \over \delta J_{A_{n}}}W.
\label{(164)}
\end{equation}
Dividing both sides of eq. (162) by $e^{iW}$ and comparing like powers 
of $\bigtriangleup J_{A}$, one obtains an infinite sequence of relations:
\begin{equation}
\langle {\underline \varphi}^{A} {\underline \varphi}^{B} \rangle
=\varphi^{A} \varphi^{B}-i G^{AB}, \;
\langle {\underline \varphi}^{A} {\underline \varphi}^{B} 
{\underline \varphi}^{C} \rangle =\varphi^{A} \varphi^{B} \varphi^{C}
-i P_{3}\varphi^{A}G^{BC}+(-i)^{2}G^{ABC}, \; {\rm etc}.,
\label{(165)}
\end{equation}
where $P$ means ``sum over the $N$ distinct permutations of indices, with
a plus sign or a minus sign according to whether the permutation of the
indices associated with fermionic fields is even or odd.'' $G^{AB}$ is 
known as the {\bf one-particle} propagator, and the
$G^{A_{1}...A_{n}}, \; n \geq 3$ are known as 
{\bf many-particle Green's functions}. They satisfy the boundary conditions
specified by the ``in'' and ``out'' states.

Any functional of the sources $J_{A}$ may be alternatively regarded as a
functional of the averages $\varphi^{A}$. From equations (163) and (164)
one sees that the one-particle propagator is the transformation matrix
from one set of variables to the other:
\begin{equation}
G^{AB}={\delta \varphi^{B} \over \delta J_{A}}.
\label{(166)}
\end{equation}
This fact may be used to establish an important relation between the 
functional $W$ and the Schwinger average of the operator field equations.
The latter is obtained from the formal functional identity
\begin{eqnarray}
0 &=& -ie^{-iW}{\overline N}[U]\int e^{i({\widetilde S}+J \varphi
+{\overline J}\chi + {\widehat J}\psi)}
{{\lvec \delta}\over \delta \varphi^{A}}
d\varphi \; d\chi \; d\psi \nonumber \\
&=& \langle {\underline {\widetilde S}}_{,A} \rangle +J_{A},
\label{(167)}
\end{eqnarray}
where ${\underline {\widetilde S}}_{,A}$ is the operator corresponding
to the functional ${\widetilde S}_{,A}$. 

If we differentiate eq. (167) on the left with respect to $J_{B}$ and
make use of (166) we obtain
\begin{equation}
G^{BC} { }_{C,} \langle {\underline {\widetilde S}}_{,A} \rangle
=-\delta_{\; A}^{B}.
\label{(168)}
\end{equation}
Here the functional derivative inside the brackets $\langle \; \rangle$
is with respect to the field operator ${\underline \varphi}^{A}$, and the
functional derivative outside the brackets is with respect to the field
average $\varphi^{C}$. ${ }_{C,} \langle {\underline {\widetilde S}}_{,A}
\rangle$ is seen to be the operator of which the one-particle propagator
$G^{BC}$ is the Green's function. Because of its boundary conditions 
$G^{BC}$ may be shown to be both a left Green's function, as in eq.
(168), and a right Green's function of ${ }_{A,} \langle
{\underline {\widetilde S}}_{,B} \rangle$ as well. It has the symmetry
\begin{equation}
G^{AB}=(-1)^{AB} G^{BA} 
\label{(169)}
\end{equation}
which implies
\begin{equation}
{ }_{A,} \langle {\underline {\widetilde S}}_{,B} \rangle
=(-1)^{A+B+AB} { }_{B,} \langle 
{\underline {\widetilde S}}_{,A} \rangle .
\label{(170)}
\end{equation}
But this is just the condition that there exist a functional
${\widetilde \Gamma}[\varphi,\chi,\psi,K,L,M]$ such that
\begin{equation}
{\widetilde \Gamma}_{,A}= \langle {\underline {\widetilde S}}_{,A}
\rangle .
\label{(171)}
\end{equation}

${\widetilde \Gamma}$ is known as the {\bf effective action}. It satisfies
the equations
\begin{equation}
{\widetilde \Gamma}_{,A}=-J_{A},
\label{(172)}
\end{equation}
\begin{equation}
{ }_{A,}{\widetilde \Gamma}_{,C}G^{CB} =-\delta_{A}^{\; B},
\label{(173)}
\end{equation}
and is related to the functional $W$ by a Legendre transformation:
\begin{equation}
W={\widetilde \Gamma}+J_{A}\varphi^{A}.
\label{(174)}
\end{equation}
This relation nay be verified through differentiation with respect to
$J_{B}$ and use of eq. (172) in the form
$$
{ }_{A,}{\widetilde \Gamma}=-(-1)^{A}J_{A}.
$$
Thus
$$
{\delta W \over \delta J_{B}}={\delta \varphi^{A}\over \delta J_{B}}
\Bigr[{ }_{A,}{\widetilde \Gamma}+(-1)^{A}J_{A}\Bigr]+\varphi^{B}
=\varphi^{B},
$$
which is just eq. (163). Since ${\widetilde \Gamma}$ is determined 
only up to an arbitrary constant of integration, eq. (174) may be
regarded as fixing it. 

${\widetilde \Gamma}$ is also known as the {\bf generating functional
for proper vertices}. This stems from its relation to the many-particle
Green's functions. By differentiating eq. (173) one can relate functional
derivatives of the one-particle propagator to derivatives of 
${\widetilde \Gamma}$. These relations yield, for example,
\begin{eqnarray}
G^{ABC}&=& {\delta \over \delta J_{A}}G^{BC}=G^{AD} \;
{ }_{D,}G^{BC} \nonumber \\
&=& (-1)^{(B+C)D+(C+D)E+(D+E)F}G^{AD}G^{BE}G^{CF}
\; { }_{DEF,} {\widetilde \Gamma}.
\label{(175)}
\end{eqnarray}
If the propagators are represented by lines and the third and higher
derivatives of ${\widetilde \Gamma}$ are represented by vertices, one
easily sees that each new differentiation with respect to a source inserts
a new line in all possible ways into the previous diagram. Each Green's
function of given order is thus representable as a sum of all the possible
tree diagrams of that order.

Suppose the spatial sections of spacetime are noncompact. Then an
$S$-matrix can be introduced, connecting states defined ``at infinity''. 
The $S$-matrix is expressible in terms of the chronological products 
appearing in eq. (162). Because these products are expressible in terms
of the Green's functions (eqs. (165)), it follows that when 
${\widetilde \Gamma}$ is used, only tree diagrams are needed in the
construction of the $S$-matrix. No closed loops appear. The vertices
generated by ${\widetilde \Gamma}$ are the {\bf proper} vertices, already
containing all quantum corrections. By noting that identical tree diagrams
occur in classical perturbation theory, but with ${\widetilde \Gamma}$ 
replaced by ${\widetilde S}$, one can show that ${\widetilde \Gamma}$ 
describes the quantum-corrected dynamics of coherent large-amplitude
fields. One must expect the same to be true also when the spatial sections
are compact and there is no $S$-matrix.

\section{The Ward-Takahashi identity}

We now resume use of the symbols $\varphi^{i},\chi_{\alpha},\psi^{\alpha},
J_{i},{\overline J}^{\alpha},{\widehat J}_{\alpha}$ and rewrite eq. (174)
in the more explicit form
\begin{equation}
W[J,{\overline J},{\widehat J},K,L,M]
={\widetilde \Gamma}[\varphi,\chi,\psi,K,L,M]
+J_{i}\varphi^{i}+{\overline J}^{\alpha}\chi_{\alpha}
+{\widehat J}_{\alpha}\psi^{\alpha}.
\label{(176)}
\end{equation}
The averages $\varphi^{i},\chi_{\alpha},\psi^{\alpha}$ depend on all
six sources, but because $K_{i},L_{\alpha}$ and $M$ do not participate 
in the Legendre transformation one may show that
\begin{equation}
{\delta W \over \delta K_{i}}={\delta {\widetilde \Gamma}\over 
\delta K_{i}}, \;
{\delta W \over \delta L_{\alpha}}={\delta {\widetilde \Gamma} \over
\delta L_{\alpha}}, \;
{\partial W \over \partial M}
={\partial {\widetilde \Gamma} \over \partial M},
\label{(177)}
\end{equation}
where the derivatives on the right refer only to the explicit dependence
of ${\widetilde \Gamma}$ on $K_{i}, L_{\alpha}$ and $M$. This result,
combined with eq. (172) in the form
$$
{ }_{A,}{\widetilde \Gamma}=-(-1)^{A}J_{A},
$$
allows eq. (159) to be rewritten as
\begin{equation}
-(-1)^{i}{\delta {\widetilde \Gamma}\over \delta \varphi^{i}}
{\delta {\widetilde \Gamma}\over \delta K_{i}}
+{\partial {\widetilde \Gamma}\over \partial M}
-(-1)^{\alpha}{\delta {\widetilde \Gamma}\over \delta \psi^{\alpha}}
{\delta {\widetilde \Gamma}\over \delta L_{\alpha}}=0,
\label{(178)}
\end{equation}
all derivatives being left derivatives. This is the {\bf Ward-Takahashi}
identity.

The Ward-Takahashi identity has important implications for the structure
of ${\widetilde \Gamma}$. That it implies the existence of some sort of
symmetry possessed by ${\widetilde \Gamma}$ becomes obvious when one notes
that, because of its BRS invariance, ${\widetilde S}$ too satisfies the
Ward-Takahashi identity. Unfortunately, to work from eq. (178) {\bf to} the
symmetry possessed by ${\widetilde \Gamma}$ is a much harder task. In
principle one might do the following. Assume that ${\widetilde \Gamma}$ 
can be expanded as a power series in $\chi_{\alpha},\psi^{\alpha},K_{i},
L_{\alpha}$ and $M$. Such an assumption has nothing {\bf a priori} to do 
with perturbation theory since the expansion is to be carried out 
{\bf after} the functional integration (155) has been performed. It is 
based on the reasonable belief that (155) varies smoothly (at least after
appropriate renormalizations) as $K_{i},L_{\alpha},M,{\overline J}^{\alpha}$
and ${\widehat J}_{\alpha}$ (and hence $\chi_{\alpha}$ and 
$\psi^{\alpha}$) go to zero.

In determining the kinds of terms that can appear in the expansion it is
useful to introduce the notion of ``ghost number.'' If one assigns the
ghost numbers $1$ to $\psi^{\alpha}$ and ${\overline J}^{\alpha}$;
$0$ to $\varphi^{i}$ and $J_{i}$; $-1$ to 
$\chi_{\alpha},{\widehat J}_{\alpha},K_{i}$ and $M$; and $-2$ to
$L_{\alpha}$; one easily sees that the integrand in (155) and the integral
itself have total ghost number zero. Hence $W$ and ${\widetilde \Gamma}$
have total ghost number zero, and all the terms in the expansion of
${\widetilde \Gamma}$ must have this property as well. Also the expansion
can contain no terms in $M$ of higher order than the first since $M$ is an
anticommuting constant.
  
If one inserts the expansion into (178) and groups together terms of like
powers, one obtains an infinite sequence of subsidiary Ward-Takahashi
identities relating the $\varphi$-dependent coefficients. Unfortunately
there seems to be no easy way of drawing simple inferences from these
identities {\bf en gros}. So far (178) has been applied only to 
renormalizable models in perturbation theory. There it has proved to be
of great service in the practical details of the renormalization program,
as well as in the demonstration that the theory is indeed renormalizable
to all orders and that unitarity is maintained. (The student is referred
to the literature for details.) What role it is destined to play in quantum
gravity remains to be seen.

\section{References in the DeWitt lectures}

\leftline{Becchi, C., A. Rouet and R. Stora (1975). 
Comm. Math. Phys. {\bf 42}, 127.}
\vskip 0.3cm
\noindent
Berezin, F.A. (1966), {\bf The Method of Second Quantization}, Academic Press,
New York.
\vskip 0.3cm
\noindent
DeWitt, B.S. (1965), {\bf Dynamical Theory of Groups and Fields}, 
Gordon and Breach, New York.
\vskip 0.3cm
\noindent
DeWitt, B.S. (1967a), Phys. Rev. {\bf 160}, 1113.
\vskip 0.3cm
\noindent
DeWitt, B.S. (1967b), Phys. Rev. {\bf 162}, 1195.
\vskip 0.3cm
\noindent
DeWitt, B.S. (1967c), Phys. Rev. {\bf 162}, 1239.
\vskip 0.3cm
\noindent
Fadde'ev, L.D. (1976), in {\bf Methods in Field Theory}, R. Balian and
J. Zinn--Justin, eds., 1975 Les Houches Lectures, North Holland, Amsterdam.
\vskip 0.3cm
\noindent
Feynman, R.P. (1948), Rev. Mod. Phys. {\bf 20}, 367.
\vskip 0.3cm
\noindent
Feynman, R.P. (1950), Phys. Rev. {\bf 80}, 440.
\vskip 0.3cm
\noindent
Feynman, R.P. (1963), Acta Phys. Polon. {\bf 24}, 697. See also in 
{\bf Proceedings of the 1962 Warsaw Conference on the Theory of Gravitation},
PWN-Editions Scientifiques de Pologne, Warszawa (1964).
\vskip 0.3cm
\noindent
Fischer, A.E. (1970), in {\bf Relativity: Proceedings of the Relativity 
Conference in the Midwest}, Carmeli, Fickler and Witten, eds., Plenum,
New York.
\vskip 0.3cm
\noindent
Freifeld, C. (1968), in {\bf Battelle Rencontres, 1967 Lectures in
Mathematics and Physics}, C.M. DeWitt and J.A. Wheeler eds., W.A. Benjamin, 
Inc., New York.
\vskip 0.3cm
\noindent
Gribov, V.N. (1977), Lecture at the Twelfth Winter School of the Leningrad
Nuclear Physics Institute. (SLAC translation No. 176).
\vskip 0.3cm
\noindent
Isham, C.J., J. Strathdee and A. Salam (1971). Phys. Rev. {\bf D3}, 1805.
\vskip 0.3cm
\noindent
Isham, C.J., J. Strathdee and A. Salam (1972). Phys. Rev. {\bf D5}, 2584.
\vskip 0.3cm
\noindent
Kostant, B. (1977), in {\bf Differential Geometrical Methods in Mathematical 
Physics} (Proceedings of the July 1-4, 1975 Symposium in Bonn), Bleuler
and Reetz, eds. Lecture Notes in Mathematics No. 570, Springer, Berlin.
\vskip 0.3cm
\noindent
Lee, B.W. (1976), in {\bf Methods in Field Theory}, R. Balian and J.
Zinn-Justin, eds., 1975 Les Houches Lectures, North Holland, Amsterdam.
\vskip 0.3cm
\noindent
Misner, C.W., K.S. Thorne and J.A. Wheeler (1973). {\bf Gravitation},
Freeman, San Francisco.
\vskip 0.3cm
\noindent
Nath, P. (1976), in {\bf Proceedings of the Conference on Gauge Theories
and Modern Field Theory}, Boston, 1975 (M.I.T. Press, Cambridge,
Massachusetts).
\vskip 0.3cm
\noindent
Stelle, K.S. (1977), Phys. Rev. {\bf D16}, 953.
\vskip 0.3cm
\noindent
Utiyama, R. (1956), Phys. Rev. {\bf 101}, 1597.
\vskip 0.3cm
\noindent
Weinberg, S. (1965), Phys. Rev. {\bf 140}, B516.
\vskip 0.3cm
\noindent
Weinberg, S. (1979), in {\bf General Relativity, an Einstein 
Centenary Survey}, S.W. Hawking and W. Israel, eds., 
Cambridge University Press.

\section{Old unification}

We can now outline how the space-of-histories formulation of sections 
2-16 provides a common ground for describing the `old' and `new' 
unifications of fundamental theories.
Quantum field theory begins once an action functional $S$ is given, since
the first and most fundamental assumption of quantum theory is that every
isolated dynamical system can be described by a characteristic action
functional \cite{DeWi65}.  The Feynman approach makes it necessary to
consider an infinite-dimensional manifold such as the space $\Phi$ of all
field histories $\varphi^{i}$.
On this space there exist (in the case of gauge theories) vector fields
\begin{equation}
Q_{\alpha}=Q_{\; \alpha}^{i} {\delta \over \delta \varphi^{i}}
\label{(179)}
\end{equation}
that leave the action invariant, i.e. (see eq. (108)), 
\begin{equation}
Q_{\alpha}S=0.
\label{(180)}
\end{equation}
The Lie brackets of these vector fields lead to a classification
of all gauge theories known so far.

\subsection{Type-I gauge theories}

The peculiar property of type-I gauge theories is that the Lie
brackets $[Q_{\alpha},Q_{\beta}]$  
are equal to linear combinations of the vector fields themselves,
with structure constants, i.e., 
\begin{equation}
[Q_{\alpha},Q_{\beta}]=C_{\; \alpha \beta}^{\gamma} \; Q_{\gamma},
\label{(181)}
\end{equation}
where $C_{\; \alpha \beta ,i}^{\gamma}=0$. The Maxwell, Yang--Mills,
Einstein theories are all example of type-I theories (this is the
`unifying feature'). All of them, being gauge theories, need
supplementary conditions, since the second functional derivative of
$S$ is not an invertible operator. After imposing such conditions, the
theories are ruled by a differential operator of D'Alembert type (or
Laplace type, if one deals instead with Euclidean field theory), or a
non-minimal operator at the very worst (for arbitrary choices of
gauge parameters). For example, when Maxwell theory is quantized via
functional integrals in the Lorenz \cite{PHMAA-34-287} 
gauge\footnote{In \cite{PHMAA-34-287} the author, L. Lorenz,
built a set of retarded potentials which can be shown to satisfy 
the Lorenz gauge, although in 1867 no-one had thought of
electrodynamics as a gauge theory. This author is not H. Lorentz, whose name is incorrectly associated to such a gauge in previous literature.}, one deals
with a gauge-fixing functional
\begin{equation}
\Phi(A)=\nabla^{b}A_{b},
\label{(182)}
\end{equation}
and the second-order differential operator acting on the potential
reads as
\begin{equation}
P_{a}^{\; b}=-\delta_{a}^{\; b}\cstok{\ }+R_{a}^{\; \; b}
+\left(1-{1\over \alpha}\right)\nabla_{a}\nabla^{b},
\label{(183)}
\end{equation}
where $\alpha$ is an arbitrary gauge parameter. The Feynman choice
$\alpha=1$ leads to the minimal operator
$$
{\widetilde P}_{a}^{\; b}=-\delta_{a}^{\; b}\cstok{\ }+R_{a}^{\; b},
$$
which is the standard wave operator on vectors in curved spacetime.
Such operators play a leading role in the one-loop expansion of the
Euclidean effective action. 

\subsection{Type-II gauge theories}

For type-II gauge theories, Lie brackets of vector fields $Q_{\alpha}$ are
as in eq. (181) for type-I theories, but the structure constants are promoted
to structure functions. An example is given by simple supergravity (a
supersymmetric gauge theory of gravity, with a symmetry
relating bosonic and fermionic fields) in four spacetime dimensions, with
auxiliary fields \cite{PRPLC-68-189}.

\subsection{Type-III gauge theories}

In this case, the Lie bracket (181) is generalized by
\begin{equation}
[Q_{\alpha},Q_{\beta}]=C_{\; \alpha \beta}^{\gamma} \; Q_{\gamma}
+U_{\; \alpha \beta}^{i} \; S_{,i},
\label{(184)}
\end{equation}
and it therefore reduces to (181) only on the {\it mass-shell}, i.e.,
for those field configurations satisfying the Euler--Lagrange equations.
An example is given by theories with gravitons and gravitinos such as
Bose--Fermi supermultiplets of both simple and extended supergravity
in any number of spacetime dimensions, without auxiliary
fields \cite{PRPLC-68-189}.

\subsection{From supergravity to general relativity}

It should be stressed that general relativity is naturally related to
supersymmetry, since the requirement of gauge-invariant Rarita--Schwinger
equations \cite{PHRVA-60-61} implies Ricci-flatness in four dimensions
\cite{PHLTA-B62-335}, which is then equivalent to vacuum Einstein equations. 
The Dirac operator is more fundamental in this framework, since
the $m$-dimensional spacetime metric is entirely re-constructed from the
$\gamma$-matrices, in that \begin{equation} g^{ab}={1\over 2m} {\rm tr}
(\gamma^{a}\gamma^{b}+\gamma^{b}\gamma^{a}).
\label{(185)}
\end{equation}

\section{New unification}

In modern high energy physics, the emphasis is no longer on fields (sections
of vector bundles in classical field theory, operator-valued distributions in
quantum field theory), but rather on extended objects such as strings.
In string theory, particles are not described as points, but
instead as strings, i.e., one-dimensional extended objects. While a point
particle sweeps out a one-dimensional worldline, the string sweeps out a
worldsheet, i.e., a two-dimensional real surface. For a free string, the
topology of the worldsheet is a cylinder in the case of a closed string, or a
sheet for an open string. It is assumed that different elementary particles
correspond to different vibration modes of the string, in much the same way
as different minimal notes correspond to different vibrational modes of
musical string instruments. The five different string theories are
different aspects of a more fundamental theory, called
$M$-theory \cite{Beck07}.
In the latest developments, one deals with `branes', which are classical
solutions of the equations of motion of the low-energy string effective
action, that correspond to new non-perturbative states of string theory,
break half of the supersymmetry, and are required by T-duality in theories
with open strings. They have the peculiar property that open strings have
their end-points attached to them.  With the language
of pseudo-Riemannian geometry, branes are timelike surfaces embedded into
bulk spacetime \cite{PHRVA-D74-084033}. According to this picture, gravity
lives on the bulk, while standard-model gauge fields are confined on the
brane. For branes, the normal vector $N$ is spacelike with respect to the
bulk metric $G_{AB}$, i.e.,
\begin{equation}
G_{AB}N^{A}N^{B}=N_{C}N^{C} >0.
\label{(186)}
\end{equation}
The action functional $S$ splits into the sum \cite{PHRVA-D74-084033}
($g_{\alpha \beta}(x)$ being the brane metric)
\begin{equation}
S=S_{4}[g_{\alpha \beta}(x)]+S_{5}[G_{AB}(X)],
\label{(187)}
\end{equation}
while the effective action \cite{DeWi03} $\Gamma$ is formally given by
\begin{equation}
e^{i \Gamma}=\int DG_{AB}(X) \; e^{i S} \times
\text{g.f. term}.
\label{(188)}
\end{equation}
In the functional integral, the gauge-fixed action reads as
(here there is summation as well as integration over repeated indices)
\begin{equation}
S_{{\rm g.f.}}=S_{4}+S_{5}+{1\over 2}F^{A}\Omega_{AB}F^{B}
+{1\over 2}\chi^{\mu}\omega_{\mu \nu}\chi^{\nu},
\label{(189)}
\end{equation}
where $F^{A}$ and $\chi^{\mu}$ are bulk and brane gauge-fixing functionals,
respectively, while $\Omega_{AB}$ and $\omega_{\mu \nu}$ are non-singular
`matrices' of gauge parameters.
The gauge-invariance properties of bulk and brane action functionals
can be expressed by saying that there exist vector fields on the
space of histories such that (cf. Eq. (180))
\begin{equation}
R_{B}S_{5}=0, \; R_{\nu}S_{4}=0,
\label{(190)}
\end{equation}
whose Lie brackets obey a relation formally analogous to Eq. (181)
for ordinary type-I theories, i.e.,
\begin{equation}
[R_{B},R_{D}]=C_{\; BD}^{A} \; R_{A}, \;
[R_{\mu},R_{\nu}]=C_{\; \mu \nu}^{\lambda} \; R_{\lambda}.
\label{(191)}
\end{equation}
The bulk and brane ghost operators are therefore
\begin{equation}
Q_{\; B}^{A}=R_{B}F^{A}=F_{\; ,a}^{A} \; R_{\; B}^{a},
\label{(192)}
\end{equation}
\begin{equation}
J_{\; \nu}^{\mu}=R_{\nu}\chi^{\mu}=\chi_{\; ,i}^{\mu} \;
R_{\; \nu}^{i},
\label{(193)}
\end{equation}
respectively. The full bulk integration means integrating first with
respect to all bulk metrics $G_{AB}$ inducing on the boundary
${\partial M}$ the given brane metric $g_{\alpha \beta}(x)$, and then
integrating with respect to all brane metrics.
Thus, one first evaluates the cosmological wave function of the bulk
spacetime \cite{PHRVA-D74-084033}, i.e.,
\begin{equation}
\psi_{\rm Bulk}=\int_{G_{AB}[\partial M]=g_{\alpha \beta}}
\mu(G_{AB},S_{C},T^{D})e^{i {\widetilde S}_{5}},
\label{(194)}
\end{equation}
where $\mu$ is taken to be a suitable measure, the $S_{C},T^{D}$ are
ghost fields, and (of course, $S_{A}$ here differs from the symbol
for the action in eq. (103))
\begin{equation}
{\widetilde S}_{5} \equiv S_{5}[G_{AB}]
+{1\over 2}F^{A}\Omega_{AB}F^{B}+S_{A}Q_{\; B}^{A} T^{B}.
\label{(195)}
\end{equation}
Eventually, the effective action results from
\begin{equation}
e^{i \Gamma}=\int
{\widetilde \mu}(g_{\alpha \beta},\rho_{\gamma},\sigma^{\delta})
e^{i {\widetilde S}_{4}}\psi_{\rm Bulk},
\label{(196)}
\end{equation}
where ${\widetilde \mu}$ is another putative measure, $\rho_{\gamma}$ and
$\sigma^{\delta}$ are brane ghost fields, and
\begin{equation}
{\widetilde S}_{4} \equiv S_{4}+{1\over 2}\chi^{\mu}\omega_{\mu \nu}
\chi^{\nu}+\rho_{\mu}J_{\; \nu}^{\mu} \sigma^{\nu}.
\label{(197)}
\end{equation}

\section{Bulk and brane BRST transformations}

This scheme is invariant under infinitesimal BRST transformations on
the bulk, given by
\begin{equation}
\delta G^{a}=R_{\; A}^{a} \; T^{A} \; \delta \Lambda,
\label{(198)}
\end{equation}
\begin{equation}
\delta S_{A}=\Omega_{AB} F^{B} \; \delta \Lambda,
\label{(199)}
\end{equation}
\begin{equation}
\delta T^{A}=-{1\over 2}C_{\; BD}^{A} T^{B}T^{D} \; \delta \Lambda,
\label{(200)}
\end{equation}
where $T^{A} \; \delta \Lambda=-\delta \Lambda \; T^{A}, \;
T^{A}T^{B}=-T^{B}T^{A}$, as well as under formally analogous BRST
transformations on the brane, i.e.
\begin{equation}
\delta g^{i}=R_{\; \mu}^{i} \; \sigma^{\mu} \; \delta \lambda,
\label{(201)}
\end{equation}
\begin{equation}
\delta \rho_{\mu}=\omega_{\mu \nu}\chi^{\nu} \; \delta \lambda,
\label{(202)}
\end{equation}
\begin{equation}
\delta \sigma^{\mu}=-{1\over 2}C_{\; \nu \zeta}^{\mu}
\sigma^{\nu} \sigma^{\zeta} \; \delta \lambda,
\label{(203)}
\end{equation}
where $\sigma^{\mu} \delta \lambda=-\delta \lambda \; \sigma^{\mu}, \;
\sigma^{\nu} \sigma^{\zeta}=-\sigma^{\zeta} \sigma^{\nu}$.

\section{New perspectives in the spectral asymptotics of Euclidean
quantum gravity}

Since the early eighties there has been a substantial revival of interest
in quantum cosmology, motivated by the hope of obtaining a complete picture
of how the universe could arise and evolve
\cite{Hawk82, Vile84, Vile86, Vile88}. By complete we here mean a
theoretical description where, by virtue of the guiding principles of
physics and mathematics, both the differential equations of the theory 
and the associated boundary (and initial) conditions are fully specified. 
Even though modern theoretical cosmology deals with yet other deep issues
such as dark matter, dark energy \cite{Kame01, Capo05} 
and cosmic strings \cite{Vile05}, the effort of
formulating the appropriate boundary conditions for the quantum state
of the universe \cite{Hawk84}, 
or at least for its (one-loop) semiclassical approximation, plays
again a key role, since the universe might have had a semiclassical origin
\cite{Hawk05}, and the various orders in $\hbar$ 
in the loop expansion describe
the departure from the underlying classical dynamics.

The physical motivations of our work result 
therefore from the following active areas of research:
\vskip 0.3cm
\noindent
(i) Functional integrals and space-time approach to quantum field theory
\cite{DeWi03}.
\vskip 0.3cm
\noindent
(ii) Attempt to derive the whole set of physical laws from invariance
principles \cite{Espo00}.
\vskip 0.3cm
\noindent
(iii) How to derive the early universe evolution from quantum physics;
how to make sense of a wave function of the universe and of 
Hartle--Hawking quantum cosmology \cite{Hart83, Hawk84}.
\vskip 0.3cm
\noindent
(iv) Spectral theory and its physical applications, including functional
determinants in one-loop quantum theory and hence the first corrections 
to classical dynamics \cite{Espo94}.

The boundary conditions that we study are part of a unified scheme for
Maxwell, Yang--Mills and Quantized General Relativity at one loop, 
i.e. \cite{Avra99}
\begin{equation}
\Bigr[\pi {\mathcal A}\Bigr]_{\mathcal B}=0,
\label{(204)}
\end{equation}
\begin{equation}
\Bigr[\Phi(A)\Bigr]_{\mathcal B}=0,
\label{(205)}
\end{equation}
\begin{equation}
[\varphi]_{\mathcal B}=0.
\label{(206)}
\end{equation}
With our notation, $\pi$ is a projector acting on the gauge field 
$\mathcal A$, $\Phi$ is the gauge-fixing functional, $\varphi$ is
the full set of ghost fields \cite{Espo97}. Both
equation (204) and (205) are preserved under infinitesimal 
gauge transformations 
provided that the ghost obeys homogeneous Dirichlet conditions as 
in (206). For gravity, we choose $\Phi$ so as to have an operator $P$
of Laplace type on metric perturbations in the one-loop Euclidean theory.

\section{Eigenvalue conditions for scalar modes}

On the Euclidean 4-ball, we expand metric perturbations $h_{\mu \nu}$
in terms of scalar, transverse vector, transverse-traceless tensor
harmonics on $S^{3}$. For vector, tensor and ghost modes, boundary
conditions reduce to Dirichlet or Robin \cite{Espo05a}. 
For scalar modes, one finds eventually the eigenvalues $E=x^{2}$ from
the roots $x$ of \cite{Espo05a}
\begin{equation}
J_{n}'(x) \pm {n\over x}J_{n}(x)=0,
\label{(207)}
\end{equation}
\begin{equation}
J_{n}'(x)+\left(-{x\over 2}\pm {n\over x}\right)J_{n}(x)=0.
\label{(208)}
\end{equation}
Note that both $x$ and $-x$ solve the same equation. For example, at small
$n$ and large $x$, the roots of Eq. (208) with $+$ sign in front of
${n\over x}$ read as (here $s=0,1,...,\infty$)
\begin{equation}
x(s,n) \sim \beta(s,n) \left[1+{\gamma_{1}\over \beta^{2}(s,n)}
+{\gamma_{2}\over \beta^{4}(s,n)}+{\gamma_{3}\over \beta^{6}(s,n)}
+{\rm O}(\beta^{-8})\right],
\label{(209)}
\end{equation}
where
\begin{equation}
\beta(s,n) \equiv \pi \left(s+{n\over 2}+{3\over 4}\right),
\label{(210)}
\end{equation}
and (having defined $m \equiv 4n^{2}$)
\begin{equation}
\gamma_{1}(m) \equiv -{(m-17)\over 8},
\label{(211)}
\end{equation}
\begin{equation}
\gamma_{2}(m) \equiv -{3455\over 384}+2m^{1/2}+{67\over 192}m
-{7\over 384}m^{2},
\label{(212)}
\end{equation}
\begin{equation}
\gamma_{3}(m)={1117523\over 15360}-{115\over 4}m^{1/2}
-{5907\over 5120}m+{3\over 4}m^{3/2}+{421\over 3072}m^{2}
-{83\over 15360}m^{3},
\label{(213)}
\end{equation}
as has been found in \cite{Espo05b}.

\section{Four generalized $\zeta$-functions for scalar modes}

From Eqs. (207) and (208) we obtain the following integral representations
of the resulting $\zeta$-functions upon exploiting the Cauchy theorem and 
rotation of contour \cite{Espo05a, Espo05b}:
\begin{equation}
\zeta_{A,B}^{\pm}(s) \equiv {(\sin \pi s)\over \pi}
\sum_{n=3}^{\infty}n^{-(2s-2)}\int_{0}^{\infty}dz \; z^{-2s}
{\partial \over \partial z}\log F_{A,B}^{\pm}(zn),
\label{(214)}
\end{equation}
where (here $\beta_{+} \equiv n, \beta_{-} \equiv n+2$)
\begin{equation}
F_{A}^{\pm}(zn) \equiv z^{-\beta_{\pm}}\Bigr(znI_{n}'(zn)
\pm n I_{n}(zn)\Bigr),
\label{(215)}
\end{equation}
\begin{equation}
F_{B}^{\pm}(zn) \equiv z^{-\beta_{\pm}}\left(znI_{n}'(zn)
+\left({(zn)^{2}\over 2} \pm n \right)I_{n}(zn)\right),
\label{(216)}
\end{equation}
$I_{n}$ being the modified Bessel functions of first kind.
Regularity at the origin is easily proved in the elliptic sectors,
corresponding to $\zeta_{A}^{\pm}(s)$ and $\zeta_{B}^{-}(s)$.

\section{Regularity of $\zeta_{B}^{+}$ at $s=0$}

We now define $\tau \equiv (1+z^{2})^{-1/2}$ and consider the uniform
asymptotic expansion (away from $\tau =1$, with notation as in
\cite{Espo05a, Espo05b})
\begin{equation}
z^{\beta_{+}}
F_{B}^{+}(zn) \sim {{\rm e}^{n\eta(\tau)}\over h(n)\sqrt{\tau}}
{(1-\tau^{2})\over \tau}
\left(1+\sum_{j=1}^{\infty}{r_{j,+}(\tau)\over n^{j}}\right),
\label{(217)}
\end{equation}
the functions $r_{j,+}$ being obtained from the Olver polynomials
for the uniform asymptotic expansion of $I_{n}$ and $I_{n}'$ \cite{Olve54}.
On splitting $\int_{0}^{1}d\tau=\int_{0}^{\mu}d\tau
+\int_{\mu}^{1}d\tau$ with $\mu$ small, we get an asymptotic expansion
of the l.h.s. by writing, {\it in the first interval} on the r.h.s.,  
\begin{equation}
\log \left(1+\sum_{j=1}^{\infty}{r_{j,+}(\tau)\over n^{j}}\right)
\sim \sum_{j=1}^{\infty}{R_{j,+}(\tau)\over n^{j}},
\label{(218)}
\end{equation}
and then computing
\begin{equation}
C_{j}(\tau) \equiv {\partial R_{j,+}\over \partial \tau}
=(1-\tau)^{-j-1}\sum_{a=j-1}^{4j}K_{a}^{(j)}\tau^{a}.
\label{(219)}
\end{equation}
The integral $\int_{\mu}^{1}d\tau$ is instead found to yield a vanishing
contribution in the $\mu \rightarrow 1$ limit \cite{Espo05b}.
Remarkably, by virtue of the spectral identity
\begin{equation}
g(j) \equiv \sum_{a=j}^{4j}{\Gamma(a+1)\over \Gamma(a-j+1)}
K_{a}^{(j)}=0,
\label{(220)}
\end{equation}
which holds $\forall j=1,...,\infty$, we find
\begin{equation}
\lim_{s \to 0}s \zeta_{B}^{+}(s)={1\over 6}\sum_{a=3}^{12}
a(a-1)(a-2)K_{a}^{(3)}=0,
\label{(221)}
\end{equation}
and 
\begin{equation}
\zeta_{B}^{+}(0)={5\over 4}+{1079\over 240}-{1\over 2}
\sum_{a=2}^{12}\omega(a)K_{a}^{(3)} 
+ \sum_{j=1}^{\infty}f(j)g(j)={296 \over 45},
\label{(222)}
\end{equation}
where
\begin{eqnarray}
\omega(a) & \equiv & {1\over 6}{\Gamma(a+1)\over \Gamma(a-2)}
\biggr[-\log(2)-{(6a^{2}-9a+1)\over 4}{\Gamma(a-2)\over \Gamma(a+1)}
\nonumber \\
&+& 2\psi(a+1)-\psi(a-2)-\psi(4)\biggr],
\label{(223)}
\end{eqnarray}
\begin{equation}
f(j) \equiv {(-1)^{j}\over j!}\Bigr[-1-2^{2-j}+\zeta_{R}(j-2)
(1-\delta_{j,3})+\gamma \delta_{j,3}\Bigr].
\label{(224)}
\end{equation}
The spectral cancellation (220) achieves three goals: (i) Vanishing of
$\log 2$ coefficient in Eq. (222); (ii) Vanishing of 
$\sum_{j=1}^{\infty}f(j)g(j)$ in Eq. (222); (iii) Regularity at the
origin of $\zeta_{B}^{+}$.

To cross-check our analysis, we evaluate $r_{j,+}(\tau)-r_{j,-}(\tau)$
and hence obtain $R_{j,+}(\tau)-R_{j,-}(\tau)$ for all $j$. Only $j=3$
contributes to $\zeta_{B}^{\pm}(0)$, and we find
\begin{eqnarray}
\zeta_{B}^{+}(0)&=& \zeta_{B}^{-}(0)-{1\over 24}\sum_{l=1}^{4}
{\Gamma(l+1)\over \Gamma(l-2)}\left[\psi(l+2)-{1\over (l+1)}\right]
\kappa_{2l+1}^{(3)} \nonumber \\
&=& {206 \over 45}+2={296\over 45},
\label{(225)}
\end{eqnarray}
in agreement with Eq. (222), where $\kappa_{2l+1}^{(3)}$ are the four
coefficients on the right-hand side of
\begin{equation}
{\partial \over \partial \tau}(R_{3,+}-R_{3,-})
=(1-\tau^{2})^{-4}\Bigr(80 \tau^{3}-24 \tau^{5}+32 \tau^{7}
-8 \tau^{9}\Bigr).
\label{(226)}
\end{equation}
Within this framework, the spectral cancellation reads as
\begin{equation}
\sum_{l=1}^{4}{\Gamma(l+1)\over \Gamma(l-2)}
\kappa_{2l+1}^{(3)}=0,
\label{(227)}
\end{equation}
which is a particular case of
\begin{equation}
\sum_{a=a_{\rm min}(j)}^{a=a_{\rm max}(j)}
{\Gamma((a+1)/2)\over \Gamma((a+1)/2 -j)}\kappa_{a}^{(j)}=0.
\label{(228)}
\end{equation}
Interestingly, the full $\zeta(0)$ value for pure gravity (i.e. including
the contribution of tensor, vector, scalar and ghost modes) is then
found to be positive: $\zeta(0)={142\over 45}$ \cite{Espo05b}, which
suggests a quantum avoidance of the cosmological singularity driven
by full diffeomorphism invariance of the boundary-value problem for
one-loop quantum theory \cite{Espo05b}.

\section{Selected open problems}

Several open problems should be brought to the attention of the reader,
and are as follows.
\vskip 0.3cm
\noindent
(i) We have encountered in sections 21-24  
a boundary-value problem where the generalized
$\zeta$-function remains well defined, even though the Mellin transform
relating $\zeta$-function to heat kernel does not exist
(see further comments below), since strong 
ellipticity is violated \cite{Avra99} (see also
\cite{Dowk97}). Are the spectral
cancellations (220) and (228) a peculiar property of the Euclidean 4-ball,
or can they be extended to more general Riemannian manifolds with
non-empty boundary?
\vskip 0.3cm
\noindent
(ii) What is the deeper underlying reason for finding
$\zeta_{B}^{+}(0)-\zeta_{B}^{-}(0)=2$? Is it possible to foresee
a geometrical or topological or group-theoretical origin of
this result?
\vskip 0.3cm
\noindent
(iii) Is it correct to say that our positive $\zeta(0)$ value for pure
gravity engenders a quantum avoidance of the cosmological singularity at
one-loop level? \cite{Espo05b, Hawk05} Does the result remain true in
higher-loop calculations or on using other regularization techniques
for the one-loop correction?
\vskip 0.3cm
\noindent
(iv) The whole scheme might be relevant for AdS/CFT in light of a 
profound link between AdS/CFT and the Hartle--Hawking wave function of
the universe \cite{Horo04}.
\vskip 0.3cm
\noindent
(v) What happens if one considers instead non-local boundary data,
e.g. those giving rise to surface states for the Laplacian? 
\cite{Espo00, Schr89, Espo99} 
\vskip 0.3cm
As far as item (i) is concerned, we should add what follows.
The integral representation (214) of the generalized $\zeta$-function
is legitimate because, for any fixed $n$, there is a countable infinity
of roots $x_{j}$ and $-x_{j}$ of eqs. (207) and (208), and they grow
approximately linearly with the integer $j$ counting such roots. The
functions $F_{A}^{\pm}$ and $F_{B}^{\pm}$ admit therefore a 
canonical-product representation \cite{Ahlf66} 
which ensures that the integral
representation (214) reproduces the standard definition of generalized
$\zeta$-function \cite{Espo05a}. Furthermore, even though the Mellin
transform relating $\zeta$-function to integrated heat kernel cannot 
be exploited when strong ellipticity is not fulfilled, it remains possible
to define a generalized $\zeta$-function. For this purpose, a weaker
assumption provides a sufficient condition, i.e. the existence of a sector
in the complex plane free of eigenvalues of the leading symbol of the
differential operator under consideration \cite{Espo05a, Espo05b}. To make
sure we have not overlooked some properties of the spectrum, we have been
looking for negative eigenvalues or zero-modes, but finding none.
Indeed, negative eigenvalues $E$ would imply purely imaginary roots
$x=iy$ of eq. (208), but such roots do not exist, as one can check both
numerically and analytically; zero-modes would be non-trivial eigenfunctions 
belonging to zero-eigenvalues, but all modes (tensor, vector, scalar 
and ghost modes) are combinations of regular Bessel functions \cite{Espo05a}
(since we require regularity at the origin of the left-hand side
of eq. (204))
for which this is impossible. As far as we can see, we still find sources
of singularities at the origin in the generalized $\zeta$-function 
resulting from lack of strong ellipticity, but the particular symmetries
of the Euclidean 4-ball background reduce them to the four terms in
Eq. (227), which add up to zero despite two of them are non-vanishing. 

In \cite{Espo05b} we
have proposed to interpret the result $\zeta(0)={142\over 45}$ for
pure gravity as an indication that {\bf full diffeomorphism invariance of
the boundary-value problem engenders a quantum avoidance of the 
cosmological singularity}. Indeed, on the one hand, the work by
Schleich \cite{Schl85} had found that, on restricting the functional
integral to transverse-traceless perturbations, the one-loop 
semiclassical approximation to the wave function of the universe diverges 
at small volumes, at least for the boundary geometry of a three-sphere.
The divergence of the wave functional does not imply, by itself, that
the probability density of the wave functional diverges at small volumes,
since the probability density $p[h]$ on the space of wave functionals
$\psi[h]$ is given by $p[h]=m[h]|\psi|^{2}[h]$, where $m[h]$ is the 
measure on this space, the scaling of which is not known in general. 
On the other hand, in our manifestly covariant evaluation of the one-loop
functional integral for the wave function of the universe, it seems 
incorrect to assume that the measure $m[h]$ scales in such a way
as to cancel exactly the contribution of the squared modulus of $\psi$,
which is proportional to the three-sphere radius raised to the power
$2\zeta(0)$. Thus, we find that our one-loop wave function of the 
universe vanishes at small volume. The normalizability condition of
the wave function in the limit of small three-geometry, which is
weaker than requiring it should vanish in this limit, was instead 
formulated and studied in \cite{Barv90}.

The years to come will hopefully tell us whether our calculations may
be viewed as a first step towards finding under which conditions a
quantum theory of gravity is singularity free in cosmology \cite{Kief05}.
For this purpose, it might also be interesting to study 
diffeomorphism-invariant boundary conditions for $f(R)$ theories of
gravity, recently studied at one-loop level on manifolds without boundary
\cite{Cogn05}.

On the non-perturbative side, encouraging progress has been made
towards finding cosmological applications of non-perturbative quantum
gravity via renormalization-group methods \cite{PHRVA-D57-971}, 
including, in particular,
theoretical models that might account for the accelerated expansion of the
universe \cite{CQGRD-23-3103} and for flat rotation curves of galaxies
\cite{CQGRD-24}.

\acknowledgments G. Esposito is grateful to the Dipartimento di
Scienze Fisiche of Federico II University, Naples, for hospitality
and support, and to Professor C. DeWitt--Morette for warm
approval of the present project. 
His work has been partially supported by PRIN {\it SINTESI}.
Springer has kindly granted permission to reprint, from section 2 to
17, the 1978 Carg\`ese Lectures by Professor B.S. DeWitt. Along the
years, G. Esposito has learned a lot from many colleagues, e.g.
I.G. Avramidi, P.D. D'Eath, A.Yu. Kamenshchik, K. Kirsten, and the
interaction with many Ph.D. students has been very helpful in the course
of preparing this teaching-oriented review.

\begin{appendix}
\section{Lie groups}
A Lie group is a group $G$ which is also a manifold with a $C^{\infty}$
structure such that the maps
$$
(x,y) \rightarrow xy
$$
$$
x \rightarrow x^{-1}
$$
are $C^{\infty}$ functions. It is indeed enough to assume that 
$(x,y) \rightarrow xy^{-1}$, or $(x,y) \rightarrow xy$, are $C^{\infty}$.
Relevant examples of Lie groups are as follows \cite{Spiv70}.
\vskip 0.3cm
\noindent
(1) The space ${\bf R}^{n}$ endowed with the addition $+$.
\vskip 0.3cm
\noindent
(2) The circle $S^{1}$ defined as the quotient space ${\bf R}/Z$.
\vskip 0.3cm
\noindent
(3) If $G$ and $H$ are Lie groups, their product $G \times H$ is also
a Lie group.
\vskip 0.3cm
\noindent
(4) The torus $S^{1} \times S^{1}$.
\vskip 0.3cm
\noindent
(5) The general linear group $GL(n,{\bf R})$, i.e. the group of all
$n \times n$ nonsingular real matrices. This is a subset of
${\bf R}^{n^{2}}$.
\vskip 0.3cm
\noindent
(6) The orthogonal group 
\begin{equation}
O(n) \equiv \left \{ A \in GL(n,{\bf R}): \; AA^{t}=I \right \}.
\label{(A1)}
\end{equation}
Here, with respect to the usual base of ${\bf R}^{n}$, the matrix $A$
represents a linear map which is an isometry, i.e. it preserves the norm
and the inner product.
\vskip 0.3cm
\noindent
(7) If $H \subset G$ is a subgroup of $G$ and also a submanifold of $G$, then
$H$ is a Lie group. Thus, for example, $S^{1} \subset {\bf R}^{2}$ is a
Lie group because 
\begin{equation}
S^{1}= \left \{ z \in {\bf C}: \; z {\overline z}=x^{2}+y^{2}=1
\right \}.
\label{(A2)}
\end{equation}
\vskip 0.3cm
\noindent
(8) The three-sphere $S^{3}$ is the Lie group of unit norm quaternions
(among all the $S^{n}$, only $S^{1}$ and $S^{3}$ admit a Lie-group
structure). On introducing the three symbols $i,j,k$ such that
\begin{equation}
i^{2}=j^{2}=k^{2}=-1,
\label{(A3)}
\end{equation}
\begin{equation}
ij=-ji=k, \; jk=-kj=i, \; ki=-ik=j,
\label{(A4)}
\end{equation}
a quaternion can be expressed in the form
\begin{equation}
x=x_{1}+x_{2}i+x_{3}j+x_{4}k,
\label{(A5)}
\end{equation}
with $(x_{1},x_{2},x_{3},x_{4}) \in {\bf R}^{4}$, so that the
complex conjugate quaternion reads as
\begin{equation}
{\overline x}=x_{1}-x_{2}i-x_{3}j-x_{4}k,
\label{(A6)}
\end{equation}
and the equation defining $S^{3}$ can be indeed satisfied, i.e.
\begin{equation}
x{\overline x}=\sum_{i=1}^{4}x_{i}^{2}=1.
\label{(A7)}
\end{equation}
\vskip 0.3cm
\noindent
(9) The group $SO(n)$ of all orthogonal matrices with unit determinant,
i.e.
\begin{equation}
SO(n) \equiv \left \{ A \in O(n): {\rm det}A=1 \right \},
\label{(A8)}
\end{equation}
is a Lie group.
\vskip 0.3cm
\noindent
(10) The group $E(n)$ of all isometries of ${\bf R}^{n}$ is a Lie
group. Every element of $E(n)$ can be written uniquely as
$A \cdot \tau$ where $A \subset O(n)$, and $\tau$ is a translation
\begin{equation}
\tau(x)=\tau_{a}(x)=x+a.
\label{(A9)}
\end{equation}
Note that $E(n) \not = O(n) \times {\bf R}^{n}$ because translations and
orthogonal transformations do not commute. It is however true that $O(n)$
is diffeomorphic to $O(n) \times {\bf R}^{n}$.

Given a Lie group $G$, its Lie algebra $g$ is the tangent space to $G$
at its identity element: $g=T_{e}G$. In general, a Lie algebra is a
finite-dimensional vector space $V$, endowed with an antisymmetric
bilinear map $[\; , \;]$ 
\begin{equation}
[X,Y]=-[Y,X] \; \; \; \; \forall X,Y \in V
\label{(A10)}
\end{equation}
which satisfies the Jacobi identity
\begin{equation}
[[X,Y],Z]+[[Y,Z],X]+[[Z,X],Y]=0 \; \; \; \; \forall X,Y,Z \in V.
\label{(A11)}
\end{equation}
An important theorem asserts that, in the finite-dimensional case,
a Lie algebra is always isomorphic to the Lie algebra $g$ of a 
finite-dimensional Lie group $G$. With a standard notation, one writes
$gl(n,{\bf R})$ for the Lie algebra of $GL(n,{\bf R})$ and 
$o(n)$ for the Lie algebra of $O(n)$. Every $A \in g$ is defined by its
value at the unit element of $G$. Given a basis $\left \{ e_{i}
\right \}$ in $g$, one has the Lie-bracket relations (see eq. (14))
\begin{equation}
[e_{i},e_{j}]=f_{\; ij}^{k} \; e_{k},
\label{(A12)}
\end{equation}
where $f_{\; ij}^{k}$ are the structure constants of $G$.
   
In quantum gravity and quantum Yang-Mills theories one deals however
with infinite-dimensional Lie groups (also called pseudo-groups in
the literature). The adjoint representation of the diffeomorphism
group is provided by a contravariant vector field $X^{\mu}$, as can
be seen from the transformation law \cite{DeWi84}
\begin{equation}
\delta X^{\mu}=\int d^{n}x' \int d^{n}x'' \; 
C_{\; \nu' \sigma''}^{\mu} X^{\sigma''}\; \delta \xi^{\nu'}
=-X_{\; ,\tau}^{\mu} \delta \xi^{\tau}+X^{\tau}
\delta \xi_{\; ,\tau}^{\mu}.
\label{(A13)}
\end{equation}
The coadjoint representation is instead provided by a covariant vector
density of unit weight according to \cite{DeWi84}
\begin{equation}
\delta Y_{\mu}=-\int d^{n}x' \int d^{n}x'' \; Y_{\sigma''}
C_{\; \; \nu' \mu}^{\sigma''} \delta \xi^{\nu'}
=-(Y_{\mu}\delta \xi^{\tau})_{,\tau}
-Y_{\sigma} \delta \xi_{\; ,\mu}^{\sigma}.
\label{(A14)}
\end{equation}
We refer the reader to the work in \cite{PHLTA-B286-251} for recent
results on gravitation as gauge theory of the diffeomorphism group,
while deformations of diffeomorphisms are studied in detail in
\cite{CQGRD-22-3511, CQGRD-23-1883}.
\end{appendix}

\end{document}